%% file: log_2026.tex
\documentclass{article}
\usepackage{hyphenat}
\usepackage{silence}
\usepackage[preprint]{log_2026}			% for preprint version
\usepackage{url}
\usepackage{makecell}
\usepackage[utf8]{inputenc}
\usepackage{amsmath}   % required by \text{...} used in the body
\usepackage{amssymb}
\usepackage{amsthm}
\usepackage{booktabs}
\usepackage{float}

\usepackage{booktabs}						% professional-quality tables
\usepackage{multirow}
\usepackage{tabularx}
\usepackage{array}
\usepackage{amsfonts}						% blackboard math symbols
\usepackage{float} % for [H] float placement
\usepackage{graphicx}						% figures
\usepackage{amsthm} % for theorem environments like definition
\usepackage{amsthm} % for theorem-like environments such as definition
\newcommand{\code}[1]{\texttt{#1}} % inline code macro
\usepackage{textgreek} % for text-mode Greek letters					% sample images
\newcommand{\smcn}{\texttt{smcn}}

\usepackage{enumitem} % for custom list spacing and margins
\usepackage{placeins}
\usepackage[numbers,compress,sort]{natbib}	% for numerical citations
\usepackage[backref=page]{hyperref}
\usepackage{graphicx}
\usepackage{flafter}
\usepackage{placeins}
\title[Beyond Homology: Non-Abelian and Orientability Invariants Sensitivity  in Cellular Networks]{Beyond Homology: Non-Abelian and Orientability Invariants Sensitivity in Cellular Networks}

\author[R. Raj et al.]{%
Raghuwansh Raj\\
University of Luxembourg \\
\email{raghuwansh.raj@uni.lu}\And
George Panagopoulos \\
University of Luxembourg\\
\email{georgios.panagopoulos@uni.lu}\And
Jun Pang\\
University of Luxembourg\\
\email{jun.pang@uni.lu}
}

\begin{document}

\maketitle

%%%%%%%%%%%%%%%%%%%%%%%%%%%%%%%%%%%%%%%%%%%
\begin{abstract}
Topological deep learning (TDL) architectures are typically evaluated on synthetic complex pairs that differ in a target invariant. Yet, whether a model recovers
that invariant or an artifact of the chosen discretizations is seldom audited. To examine this, we design
and test pairs of complexes that look identical to standard homological and local message-passing tests, have the same Hasse graphs, yet differ in the target invariant. We study two beyond-homology regimes: non-Abelian fundamental-group structure and homotopy-matched orientability. For the former, we construct genus-2 pairs that match homology, cell statistics, local-incidence profiles, and rank-colored Hasse 1-WL behavior while differing in their fundamental groups. For the latter, we evaluate punctured torus–Klein bottle pairs and show that distance, boundary, cycle, and spectral signals can confound apparent orientability detection, motivating stronger matched controls. Under the tested settings, subcomplex-lifted architectures separate some pairs beyond the matched homological, local-incidence, and Hasse 1-WL controls, while a depth sweep reveals an empirical depth-dependent trend without establishing a general barrier or direct computation of the target invariants. These results show that claims of topological expressivity must be stated relative to the model representation, explicitly controlled information channels, and the precise scope of the benchmark.
\end{abstract}

%%%%%%%%%%%%%%%%%%%%%%%%%%%%%%%%%%%%%%%%%%%
\section{Introduction}
\label{sec:introduction}

Topology studies properties of spaces that survive continuous deformation. Topological deep learning (TDL) makes such structures available to neural networks by replacing graphs with richer domains, including simplicial, cellular, and combinatorial complexes~\citep{hajij2022tdl,papillon2023survey}. These representations contain cells of several dimensions and record how vertices, edges, faces, and higher-dimensional cells are incident. Message Passing Simplicial Networks (MPSNs)~\citep{bodnar2021simplicial} and Cellular Isomorphism Networks (CINs)~\citep{bodnar2021cellular} use these relations to propagate information beyond ordinary graph connectivity. More recently, Multi-Cellular Networks (MCNs) and their scalable variant, Scalable Multi-Cellular Networks (SMCNs), were introduced to address blind spots of conventional higher-order message passing~\citep{eitan2024blindspots}.

This progress raises a basic empirical question. \textbf{When a cellular network distinguishes two topologically different complexes, has it detected the intended invariant?} The answer is not automatic. A neural network never receives a topological space directly. It receives a finite representation containing cell counts, incidence relations, adjacency tensors, features, and possibly positional or distance encodings. Two spaces may differ in orientability, homology, or fundamental group while their discretizations also differ in a much simpler statistic. A model can then produce the correct answer without recovering the intended topology. It may count boundary components, recognize a degree pattern, detect a short cycle, or exploit a shortest-path distribution introduced by the construction. This is an instance of \emph{shortcut learning} and does not support the intended interpretation of the result~\citep{geirhos2020shortcut}.

We study this issue using two global topological invariants, orientability and the fundamental group. Orientability asks whether a surface admits a globally consistent choice of orientation. Equivalently, it can be understood as the ability to choose a coherent normal direction across the surface~\citep{hatcher2002algebraic,stillwell1993classical}. The fundamental group records loops up to homotopy and retains their noncommutative relations~\citep{hatcher2002algebraic,lyndon1977combinatorial}. Unlike homology, which provides an abelian summary of cycles, the fundamental group \(\pi_1\) can distinguish spaces with the same homology~\citep{hatcher2002algebraic,massey1991basic}.
This separates the question ``Does the model detect global topology?'' from the weaker question ``Does the model exploit how the topology was encoded?'' When revisiting standard orientability pairs such as the cylinder vs. the M\"obius band, the torus vs. the Klein bottle, and the punctured torus vs. the punctured Klein bottle, we find that a small multilayer perceptron (MLP) using only node-to-cell shortest-path histograms can separate many pairs. We call this diagnostic model the distance multilayer perceptron (\textbf{DIST\_MLP}). Success on these benchmarks therefore cannot automatically be interpreted as computing orientability. It may instead arise from a metric shortcut introduced by the gluing pattern.

This paper develops an empirical argument about the controls required before pair separation can support a topological interpretation. Our position consists of two claims.

\begin{enumerate}[leftmargin=1.6em,itemsep=1pt]

\item \textbf{Matching selected statistics does not certify the target invariant.}
Pairs matched on standard topological and combinatorial statistics may still contain unintended shortcuts. In all three orientability families studied, global metric information provides a shortcut across the tested constructions. Their separation therefore cannot, by itself, be interpreted as orientability detection.

\item \textbf{Restricted probes make alternative explanations testable.}
A probe is a diagnostic model trained on the same pair separation task as the main architectures but given access to only one predefined information channel. For example, \textbf{DIST\_MLP} receives only shortest-path histograms and performs no message passing. If this probe separates a pair, metric information alone is sufficient to explain the success of any larger model using the same channel. A failed probe rules out only the tested features, model class, and training budget.
  
\end{enumerate}

Our work makes the following contributions.

\begin{itemize}

\item \textbf{A representation-relative framework for synthetic invariant benchmarks.}
We distinguish the fact that two spaces differ topologically from evidence that a model detects this difference. We then test whether a simple restricted model can separate them using unintended information. Our framework combines eight shortcut channels, exact generator checks, empirical probes, permutation reliability controls, and explicit limits on the resulting claims.

\item \textbf{Audited orientability benchmarks and a partially controlled non-Abelian benchmark.}
We study the cylinder vs. M\"obius band, torus vs. Klein bottle, and punctured torus vs. punctured Klein bottle, and find that their separation can be explained by unintended metric or boundary information rather than orientability alone. We construct genus-2 surface-group versus free-group complexes with different fundamental groups but matched homological, combinatorial, and local structural properties. On this partially controlled family, architectures more expressive than the tested baselines bounded by one-dimensional Weisfeiler–Leman (\(1\)-WL)~\citep{xu2019powerful} or one-dimensional Cellular Weisfeiler–Leman (\(1\)-CWL)~\citep{bodnar2021cellular} achieve separation, providing certificate-qualified evidence of greater structural expressivity.

\end{itemize}

More broadly, our study shows that claims of topological expressivity must be evaluated relative to the model representation, the controlled information channels, and the precise scope of the benchmark.All analysis code, together with the scripts that reproduce every figure and table in this article and instructions for obtaining and preparing the benchmark data, is available in a public repository at \href{https://github.com/turyam4/Beyond_Homology}{github}.

\section{Related Work}
\label{sec:related}

We review the literature most closely connected to our methodology. We first discuss higher-order message passing and global structural encodings, and then relate our channel controls and pairwise evaluation protocol to research on shortcut learning and reliable graph benchmarking.

TDL extends graph message passing from graphs to simplicial, cellular, and combinatorial complexes~\cite{hajij2022tdl,papillon2023survey,hajij2020cellcomplex,hajij2024topox}.  MPSNs~\cite{bodnar2021simplicial} and  CINs~\cite{bodnar2021cellular} connect higher-order message passing with Weisfeiler–Leman refinement. After suitable lifts, these models can distinguish structures that ordinary one-dimensional Weisfeiler–Leman (\(1\)-WL) refinement cannot~\cite{xu2019powerful}. CIN++, an extension of CIN, introduces additional interactions between higher-dimensional cells~\cite{giusti2023cinpp}. Other approaches use Hodge Laplacians~\cite{ebli2020simplicial}, orientation-equivariant layers~\cite{roddenberry2021principled}, simplicial and cellular attention~\cite{battiloro2023attention,ballester2024cellular}, higher-order signal processing~\cite{barbarossa2020topological,schaub2021signal}, or cellular sheaves~\cite{hansen2019sheaf}.
These methods preserve different kinds of information. Some explicitly use signed boundary operators or orientation data, whereas the models tested in our study receive unsigned incidence relations. Orientability is therefore not supplied directly, and successful separation may depend on another property retained by the representation. Known blind spots of higher-order message passing further motivate the use of carefully controlled benchmark pairs~\cite{eitan2024blindspots}.

Global positional information can also distinguish structures that appear identical from local message passing. Graphormer uses shortest-path distance as a spatial bias~\cite{ying2021graphormer}, while spectral and learnable positional encodings provide related global information~\cite{kreuzer2021san,dwivedi2022lspe,lim2023signnet}. Such encodings are now common components of graph architectures~\cite{rampasek2022graphgps}. Our benchmark design draws on research concerning shortcut learning and controlled graph evaluation. Neural models may rely on correlations that are predictive within a dataset but unrelated to the intended task~\cite{geirhos2020shortcut}, including incidental structural properties of graphs~\cite{bechlerspeicher2024graphs}. Studies have examined whether graph benchmarks can be distinguished by \(1\)-WL~\cite{zopf2022wl}, how synthetic data generation influences experimental conclusions~\cite{yoon2023graphworld}, and how evaluation protocols affect reported performance~\cite{shchur2018pitfalls,errica2020fair,dwivedi2023benchmarking}. Existing graph datasets~\cite{morris2020tudataset,hu2020ogb} and topological learning pipelines~\cite{telyatnikov2024topobench,hajij2024topox} provide standardized evaluation settings, but additional controls are needed to identify the information responsible for model separation. We compute integral homology and torsion from cellular boundary matrices using Smith normal form~\cite{kannan1979polynomial,storjohann1996near}, and test selected representation channels using restricted trainable probes.
Moreover, our evaluation protocol builds on the pairwise testing framework of the BREC expressiveness dataset~\cite{wang2024brec}. BREC evaluates whether a model distinguishes two nonisomorphic instances while remaining stable under permutations of the same instance. We adopt this distinction between target separation and permutation reliability, using separation statistics based on Hotelling \(T^2\) and Mahalanobis distance~\cite{hotelling1931generalization,mahalanobis1936generalized}. We extend the protocol by specifying potential shortcut channels and evaluating restricted probes for these channels before interpreting more expressive models. This allows model separation to be reported relative to the information controlled in each benchmark family.

%%%%%%%%%%%%%%%%%%%%%%%%%%%%%%%%%%%%%%%%%%%
\section{Background and Preliminaries}
\label{sec:pre}
\subsection{Notation}
A finite \(d\)-dimensional combinatorial complex is denoted by
\(X=\bigsqcup_{k=0}^{d}X_k\), where \(X_k\) is the set of \(k\)-dimensional
cells, \(\bigsqcup\) denotes disjoint union, and \(|X_k|\) is the number of
such cells. The relation \(\sigma\prec\tau\) means that \(\sigma\) is a face
of \(\tau\). We use \(\operatorname{Real}(X)\) for the topological realization
of \(X\), while \(\Phi(X)\) denotes the tensor representation supplied to the
model.

The target invariant is denoted by \(\tau\), and
\(\mathcal C=\{C_1,\ldots,C_m\}\) is the set of audited information channels.
The Euler characteristic is \(\chi(X)\), the \(k\)th integral homology group is
\(H_k(X;\mathbb Z)\), and \(\beta_k=\operatorname{rank}H_k(X;\mathbb Z)\) is
its \(k\)th Betti number. Here, \(\mathbb Z\) denotes the integers,
\(\partial_k\) is the cellular boundary operator, and
\(H_*=(H_0,H_1,H_2)\). The fundamental group is denoted by \(\pi_1(X)\),
\(F_n\) is the free group on \(n\) generators, \([a,b]=aba^{-1}b^{-1}\) is a
commutator, and \(\vee\) denotes the wedge sum. For surface constructions,
\(g\), \(q\), and \(b\) denote orientable genus, crosscap number, and boundary
count, while \(h\) and \(p\) denote grid height and seam length.

%%%%%%%%%%%%%%%%%%%%%%%%%%%%%%%%%%%%%%%%%%%
\subsection{Topological Invariants}

A \textbf{topological invariant} is a quantity or algebraic object preserved under an equivalence such as homeomorphism or homotopy equivalence~\citep{hatcher2002algebraic,munkres1984elements}. We consider Euler characteristic, homology, orientability, and the fundamental group. A finite \(d\)-dimensional \textbf{combinatorial complex \((\mathrm{CC})\)} is \(X=\bigsqcup_{k=0}^{d}X_k\), where \(X_k\) is the set of \(k\)-dimensional cells and \(\sigma\prec\tau\) means that \(\sigma\) is a face of \(\tau\).
\paragraph{Euler Characteristic}
For a finite two-dimensional complex, \(\chi(X)=|X_0|-|X_1|+|X_2|\). Since it depends only on cell counts, \(\chi(X)\) provides a simple global control.

\paragraph{Homology}
The \(k\)th integral homology group is \(H_k(X;\mathbb Z)=\ker(\partial_k)/\operatorname{im}(\partial_{k+1})\), where \(\partial_k\) is the cellular boundary operator. The groups \(H_0\), \(H_1\), and \(H_2\) describe connected components, one-dimensional cycles, and two-dimensional cycles. The Betti number \(\beta_k=\operatorname{rank}H_k(X;\mathbb Z)\) records the free rank but not torsion, so we use exact integral homology when matching benchmark pairs~\citep{hatcher2002algebraic,massey1991basic,kannan1979polynomial,storjohann1996near}.
%%%%%%%%%%%%%%%%%%%%%%%%%%%%%%%%%%%%%%%%%%%
Further background on combinatorial complexes, Euler characteristic, and cellular homology is provided in Appendix~\ref{app:topological-background}.

\subsection{Presentation Complexes}
A group presentation \(P=\langle g_1,\ldots,g_n\mid r_1,\ldots,r_m\rangle\) describes a group using generators \(g_i\) and relations \(r_j\). Its presentation complex has one vertex, one loop per generator, and one face per relation, with the fundamental group \(\pi_1(K(P))\cong G(P)\)~\citep{hatcher2002algebraic,lyndon1977combinatorial}.Our Fundamental Group Family compares the genus \(2\) surface relation \([a,b][c,d]\) with a word that freely reduces to the identity. Their fundamental groups are different, but both have homology \(H_*=(\mathbb Z,\mathbb Z^4,\mathbb Z)\). The Orientability Family is built by gluing rectangular grids either in the same order or in reverse order, producing the three orientable vs. nonorientable pairs. The mathematical details of these constructions and information about the emitted tensors and the further information related to the model input are provided in Appendix~\ref{app:topological-background}.

\begin{table}[!t]
\centering
\caption{Shortcut channels audited during benchmark certification.}
\label{tab:channels}
\footnotesize
\begin{tabular}{@{}llll@{}}
\toprule
& \textbf{Channel} & \textbf{Test} & \textbf{Audit mechanism} \\
\midrule
C1 & Cell counts 
& \(|V|, |E|, |F|, \chi\) equal 
& gate + \texttt{COUNT\_MLP} \\

C2 & Homology 
& Betti over \(\mathbb{Q}, \mathbb{F}_2, \mathbb{F}_3\); integral \(H_\ast\) by SNF 
& gate \\

C3 & Degree / incidence 
& sorted vertex-degree and edge--face incidence multisets 
& gate \\

C4 & Boundary structure 
& number of boundary loops and their lengths 
& gate + \texttt{BOUNDARY\_MLP} \\

C5 & Global metric
& node--node / node--cell SPD histograms; eccentricity 
& \texttt{DIST\_MLP} \\

C6 & Spectral 
& adjacency / Laplacian spectrum 
& post-hoc audit \\

C7 & Short cycles 
& girth; cycle-count spectrum up to length \(L\) 
& post-hoc audit \\

C8 & Typed WL 
& rank-colored Hasse \(1\)-WL to stabilization 
& gate + \texttt{HASSE\_GIN} \\
\bottomrule
\end{tabular}
\end{table}

% Methodology Section
%%%%%%%%%%%%%%%%%%%%%%%%%%%%%%%%%%%%%%%%%%%
section{Methodology}
\label{sec:methodology}
A topological benchmark begins with two complexes $X$ and $Y$ whose target invariant differs, $\tau(X)\neq\tau(Y)$. This is necessary, but not sufficient. A neural model receives a finite representation of the underlying space. It may therefore separate the pair using a cheaper correlate such as boundary count, degree profile, cycle length, or shortest-path distance, without computing $\tau$. The central question is consequently not only \emph{whether the model can separate the pair,} but \emph{which input channel makes separation possible?}

\subsection{Channel Taxonomy}
Let \(X\) and \(Y\) be finite combinatorial complexes, and let \(|X|\) and \(|Y|\) denote their underlying topological spaces. The target invariant is written as \(\tau(|X|)\). A preprocessing map \(\Phi\) converts each complex into the collection of tensors supplied to the model:
\[
\Phi(X)=
\{\text{cell features, unsigned incidence, adjacency, and bridge relations}\}.
\]
Thus, the model receives \(\Phi(X)\), not the complete complex \(X\) or its topological realization \(|X|\).

Let \(\mathcal C=\{C_1,\ldots,C_m\}\) be the audited shortcut channels, where each \(C_j\) extracts a particular type of information from \(\Phi(X)\). A valid benchmark pair must differ in the target invariant,
\[
\tau(|X|)\neq\tau(|Y|),
\]
while each audited channel must either satisfy
\[
C_j(\Phi(X))=C_j(\Phi(Y))
\]
or be explicitly reported as uncontrolled.

We organize possible shortcuts into channels according to the information needed to compute them, ranging from simple counts to homology, boundary structure, local incidence, global distances, spectra, cycles, and Hasse \(1\)-WL refinements. Table~\ref{tab:channels} defines these channels.  Table~\ref{tab:combined-channel-controls} reports the status of these channels for all benchmark pairs. Each channel is assigned one status relative to the explicitly declared
tests. A channel is \emph{matched} when all declared statistics agree
between the two members of the pair. A \emph{leak} is a measured difference 
or a successful restricted probe that provides an alternative explanation
for separation. A channel is \emph{partial} when only some aspects of that
channel have been tested, and \emph{open} when the channel has not been
examined sufficiently. At the aggregate-channel level, a partial status may contain a mixture of matched and leaking audited subchannels. Probe success or failure is an experimental outcome,
not a channel status. In particular, a failed probe rules out only its tested
feature map, model class, optimization budget, and parameter range.

\subsection{Benchmark Family I Orientability}
\label{sec:orientability-family}

\begin{figure*}[t]
\centering
\caption{The three orientability benchmark pairs. Each pair contrasts an orientable construction with a nonorientable construction obtained from the corresponding grid identification.}
\label{fig:orientability-family}
\begin{minipage}[t]{0.32\textwidth}
\centering
\includegraphics[width=\linewidth]{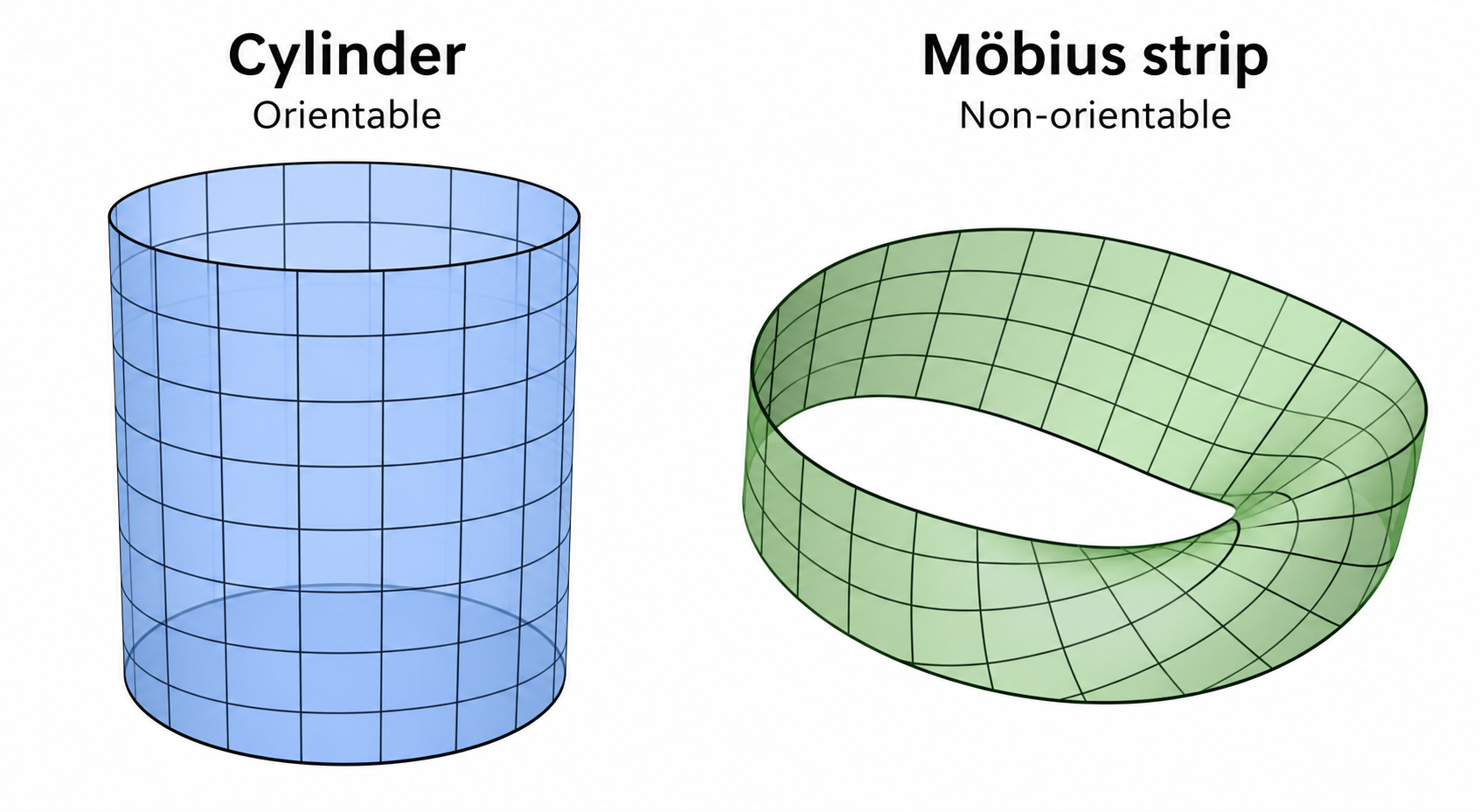}
\caption*{(a) Cylinder vs. M\"obius band}
\end{minipage}
\hfill
\begin{minipage}[t]{0.32\textwidth}
\centering
\includegraphics[width=\linewidth]{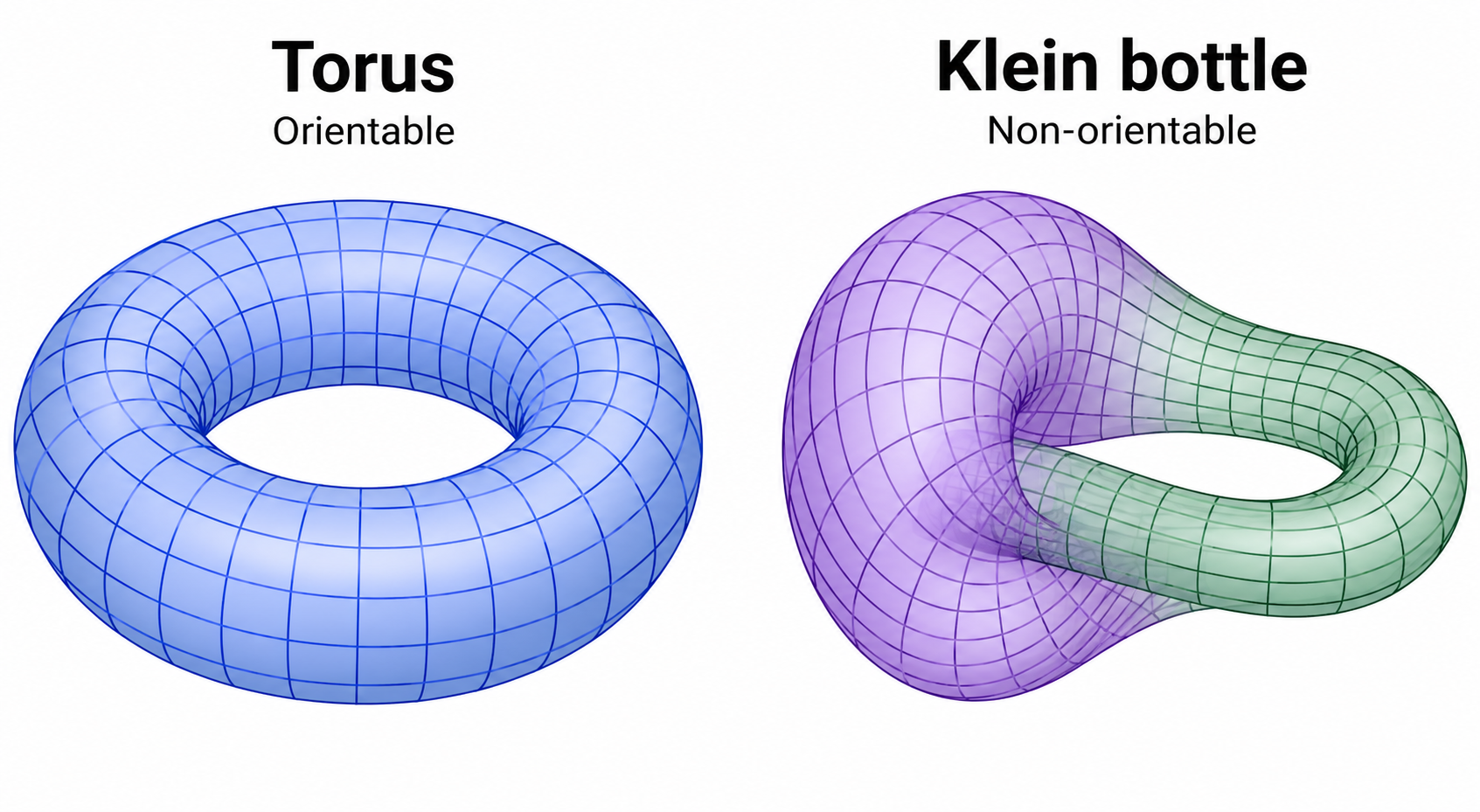}
\caption*{(b) Torus vs. Klein bottle}
\end{minipage}
\hfill
\begin{minipage}[t]{0.32\textwidth}
\centering
\includegraphics[width=\linewidth]{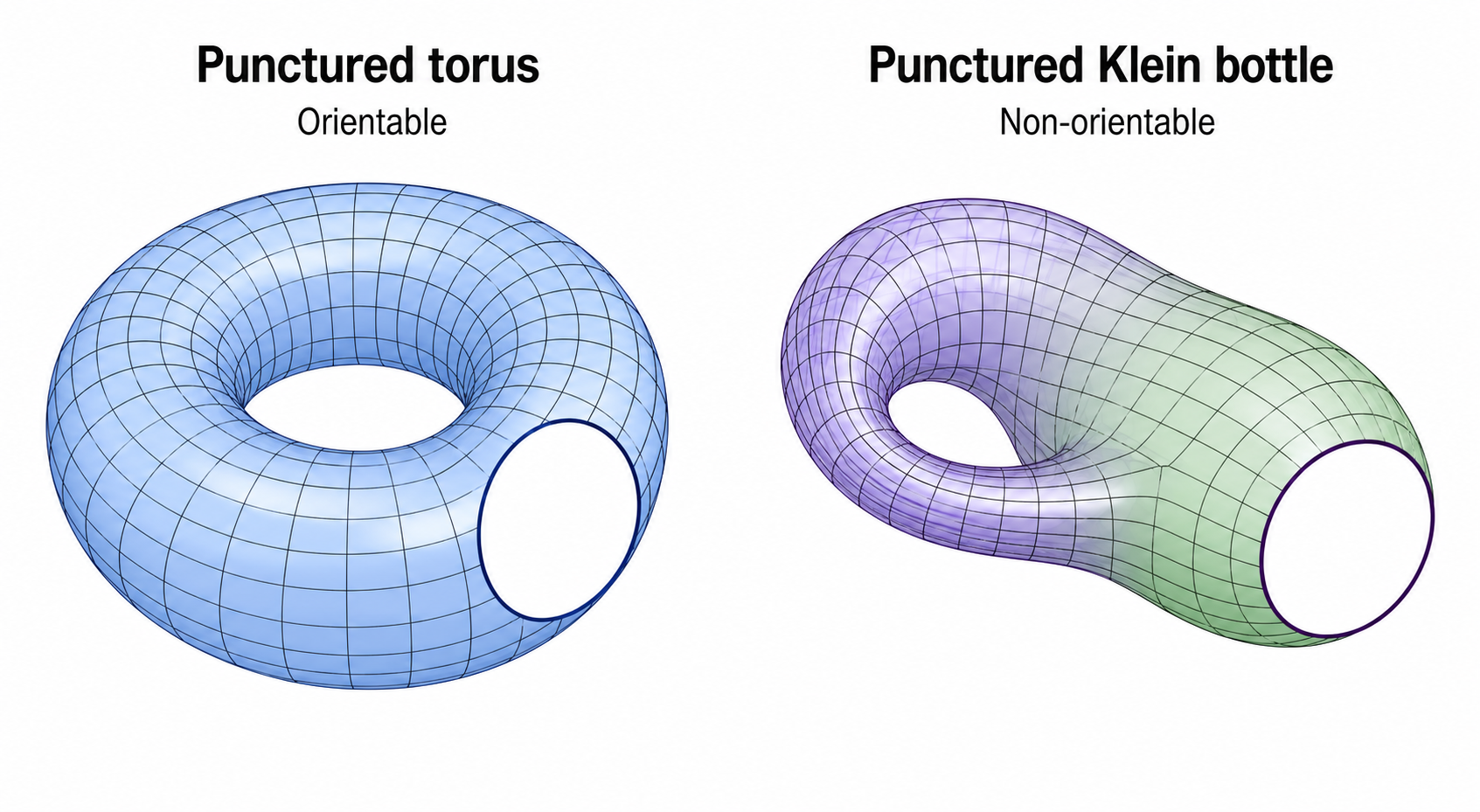}
\caption*{(c) Punctured torus vs. punctured Klein bottle}
\end{minipage}

\end{figure*}

We now apply the channel taxonomy to two benchmark families. The first family contains three orientability pairs constructed from rectangular grids (Appendix~\ref{app:orient-construction}), while the second contains a pair of presentation complexes with different fundamental groups. For every pair, we identify the intended topological difference and report each channel as matched, leaking, partially controlled, or open.
The Orientability Family contains three paired constructions obtained from rectangular cell grids.  Figure~\ref{fig:orientability-family} visualizes the three pairs in the Orientability Family. The drawings show the intended topological contrast in each case: an orientable member is paired with a nonorientable member obtained by changing the grid identification. However, the figure should not be read as evidence that the benchmark isolates orientability. The later channel table shows that these constructions also expose boundary, homology, metric, or cycle information, depending on the pair. Thus, the figure motivates the benchmark family visually, while the certification table determines what can be claimed from model separation.

\textbf{Cylinder and M\"obius Band.} The cylinder and M\"obius band differ in orientability but have the same homotopy type \(S^1\), where \(S^1\) denotes the circle. Both have homology \(H_*=(\mathbb Z,\mathbb Z,0)\), where \(\mathbb Z\) denotes the integers. Their cell counts and local incidence profiles also agree, matching channels \(C_1\) through \(C_3\).
Their boundary structures differ because the cylinder has two boundary loops of length \(p\), while the M\"obius band has one boundary loop of length \(2p\). This produces a leak through \(C_4\). Their global distance distributions also differ, producing a leak through \(C_5\). Separation of this pair, therefore, cannot be attributed specifically to orientability.

% Table 

\begin{table*}[!ht]
\centering
\caption{Channel status across the four benchmark pairs. A matched channel
agrees on all declared statistics. A leak provides a measured alternative
explanation for separation. A partial channel has only some declared
subchannels matched or audited, whereas an open channel remains insufficiently
examined. Superscript \textsuperscript{d} refers to the metric audit in
Appendix~\ref{app:c5}; \textsuperscript{e} refers to the cycle audit in
Appendix~\ref{app:mc4}; and \textsuperscript{f} refers to the typed-refinement
audit in Appendix~\ref{app:depth-sweep}.}
\label{tab:combined-channel-controls}
\scriptsize
\renewcommand{\arraystretch}{1.15}
\setlength{\tabcolsep}{3pt}

\begin{tabular}{@{}lp{7.0cm}cccc@{}}
\toprule
\textbf{Channel} &
\textbf{Evidence or diagnostic} &
\shortstack{\textbf{Cyl. vs.}\\\textbf{M\"ob.}} &
\shortstack{\textbf{Tor. vs.}\\\textbf{Klein}} &
\shortstack{\textbf{Punctured}\\\textbf{pair}} &
\shortstack{\textbf{Surface vs.}\\\textbf{free group}}\\
\midrule

\(C_1\) &
Rank-wise cell counts and Euler characteristic &
\textbf{matched} &
\textbf{matched} &
\textbf{matched} &
\textbf{matched}\\

\(C_2\) &
Exact integral homology, including torsion &
\textbf{matched} &
\textbf{leak} &
\textbf{matched} &
\textbf{matched}\\

\(C_3\) &
Local incidence and vertex-link profiles &
\textbf{matched} &
\textbf{matched} &
\textbf{matched} &
\textbf{matched}\\

\(C_4\) &
Boundary counts and boundary loop lengths &
\textbf{leak} &
\textbf{matched} &
\textbf{matched} &
\textbf{matched}\\

\(C_5\) &
Global node-to-cell shortest-path distances &
\textbf{leak} &
\textbf{leak} &
\textbf{leak} &
\textbf{partial\textsuperscript{d}}\\

\(C_6\) &
Adjacency and Laplacian spectral information &
\textbf{open} &
\textbf{open} &
\textbf{open} &
\textbf{leak}\\

\(C_7\) &
Girth, short cycle counts, and depth-visible cycles &
\textbf{open} &
\textbf{open} &
\textbf{partial} &
\textbf{partial\textsuperscript{e}}\\

\(C_8\) &
Rank-colored Hasse \(1\)-WL and refinement controls &
\textbf{matched} &
\textbf{open} &
\textbf{partial\textsuperscript{f}} &
\textbf{matched}\\

\bottomrule
\end{tabular}
\end{table*}

\textbf{Torus vs. Klein Bottle.}
The torus and Klein bottle are closed surfaces with equal cell counts, Euler characteristic \(\chi=0\), local incidence profiles, and zero boundary signal.
%where \(\chi\) denotes the Euler characteristic.
Let \(T^2\) denote the torus and \(K\) denote the Klein bottle. Their integral homology groups are
\[
\begin{aligned}
H_*(T^2)
  &= (\mathbb Z,\mathbb Z^2,\mathbb Z),\\
H_*(K)
  &= (\mathbb Z,\mathbb Z\oplus\mathbb Z_2,0).
\end{aligned}
\]
where \(\mathbb Z_2\) denotes the integers modulo \(2\). Homology therefore produces a leak through \(C_2\). Their global distance distributions also differ, producing a leak through \(C_5\). This pair is consequently not a homology-controlled orientability benchmark.

\textbf{Punctured Torus vs. Punctured Klein Bottle.}
Removing a disk from each closed surface produces two spaces with one boundary component. Both have homology \(H_*=(\mathbb Z,\mathbb Z^2,0)\) and homotopy type \(S^1\vee S^1\), where \(\vee\) denotes the one point union. Their cell counts, local incidence profiles, and boundary statistics agree, matching channels \(C_1\) through \(C_4\).

Typed Hasse \(1\)-WL does not distinguish the tested \(p=4\) subset, but it separates the tested cases with \(p\geq6\) at round \(h+2\); thus \(C_8\) is only partially controlled for the full family. However, global shortest-path distributions distinguish the pair through \(C_5\), and a depth-dependent cycle signal leaves \(C_7\) only partially controlled. This is the strongest pair in the Orientability Family, but it remains confounded by metric information.

\subsection{Benchmark Family II: Fundamental Group}
\label{sec:fundamental-group-family}

\paragraph{Genus-\(2\) Surface Group vs. Free Group}
The Fundamental Group Family compares two presentation complexes constructed from a wedge of four circles and one attached \(2\)-cell. Let \(X\) denote the surface group complex and \(Y\) denote the free group complex. In \(X\), the attaching word reduces to \([a,b][c,d]\), where \([a,b]=aba^{-1}b^{-1}\) denotes the commutator of \(a\) and \(b\). In \(Y\), the attaching word freely reduces to the identity. Further details of the presentation-complex construction are given in Appendix~\ref{app:fg-construction}.

The resulting fundamental groups are
\[
\pi_1(X)=\langle a,b,c,d\mid [a,b][c,d]\rangle,
\qquad
\pi_1(Y)=F_4,
\]
where \(\pi_1\) denotes the fundamental group and \(F_4\) denotes the free group on four generators. These groups are not isomorphic, although both complexes have homology \(H_*=(\mathbb Z,\mathbb Z^4,\mathbb Z)\). The complex \(X\) is the closed orientable genus \(2\) surface \(\Sigma_2\), while \(Y\) has homotopy type \((\bigvee^4 S^1)\vee S^2\). The main topological properties of all benchmark pairs are summarized in Appendix~\ref{app:topological-background}.

The Fundamental Group Family matches exact integral homology, rank-wise cell
counts, local-incidence and vertex-link profiles, boundary statistics,
selected word statistics, and rank-colored Hasse $1$-WL through
stabilization. Its remaining controls are incomplete. Failure of
\textsc{Dist\_MLP} excludes only its aggregated shortest-path feature map and
training budget. A post-hoc audit finds that the aggregate histogram matches on all $16$
pairs, whereas the finer per-vertex marking profile, consumed by shortest-path
SMCN, matches on only $6/16$. The tested spectra differ on all $16$ pairs, and
cycle counts diverge beyond triangles. Consequently, separation on the full
family establishes sensitivity to information beyond the matched homological,
local, and Hasse-$1$-WL controls, but not direct or causal sensitivity to
$\pi_1$. Accordingly, the global metric channel \(C_5\) is partial, the spectral
channel \(C_6\) is a leak, and the cycle channel \(C_7\) remains partial,
with detected leakage in the audited cycle statistics.

\begin{figure*}[!t]
    \centering
    \includegraphics[
        width=\textwidth,
        keepaspectratio
    ]{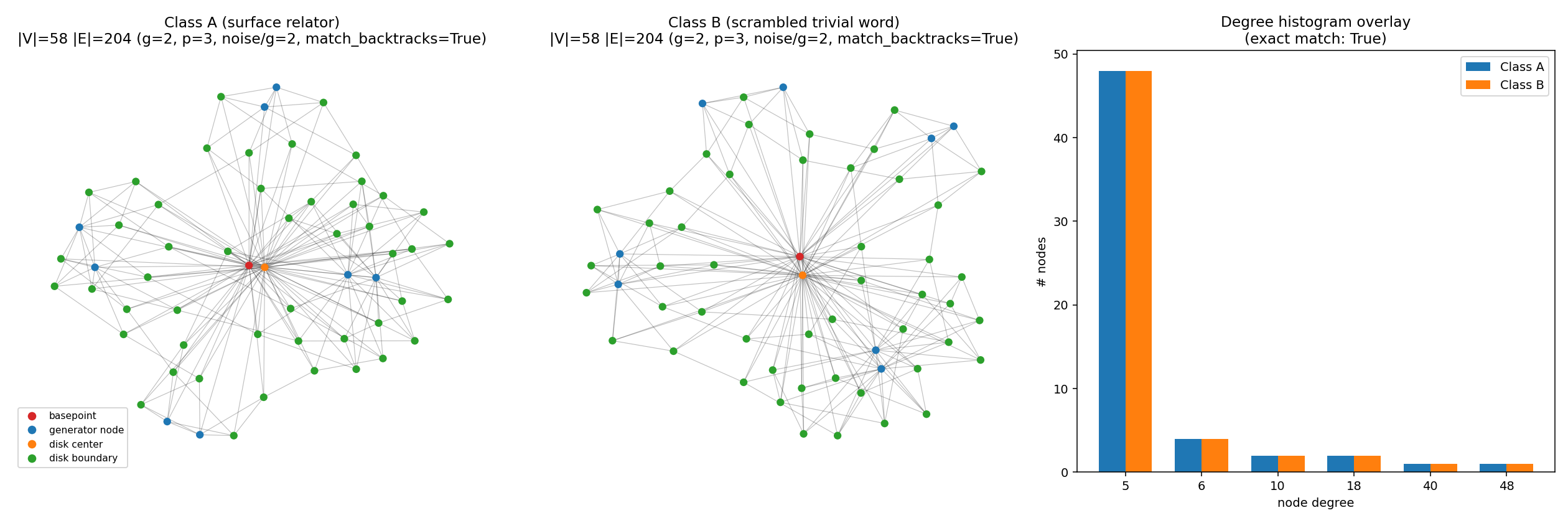}
    \caption{Example from the Fundamental Group benchmark. Class A is generated from the genus \(2\) surface relator, whereas Class B uses a freely trivial scrambled word. The matching degree histograms illustrate one of the controlled local statistics.}
    \label{fig:fg-pair-example}
\end{figure*}

% Required in preamble
% \usepackage{graphicx}

\subsection{The Baseline Ladder}
\label{sec:baseline-ladder}
The baseline ladder is ordered by information access rather than parameter count. Table~\ref{tab:baseline-ladder} summarizes the models used in this ladder. The first three models are restricted probes: \textbf{COUNT\_MLP} sees only cell counts and Euler characteristic, \textbf{BOUNDARY\_MLP} sees only boundary loop statistics, and \textbf{DIST\_MLP} sees only node-to-cell shortest-path summaries. These probes test whether a named shortcut is already sufficient for pair separation.

The remaining models receive richer structure. \textbf{GIN} uses the \(1\)-skeleton, \textbf{ESAN} uses graph substructure bags, \textbf{HASSE\_GIN} uses ranked Hasse relations, \textbf{CIN/HOMP} uses unsigned cellular message passing, and \textbf{SMCN} uses a face-marked subcomplex lift~\citep{xu2019powerful,bevilacqua2022esan,bodnar2021cellular,eitan2024blindspots}. If a restricted probe separates a pair, then success by a richer model with access to the same channel cannot be interpreted as invariant detection. Conversely, failure of a baseline rules out only the tested mechanism, not all possible shortcuts. Further implementation details for the probes and baselines are given in Appendix~\ref{app:experimental-details}.

\begin{table}[!t]
\centering
\caption{The baseline ladder. \textit{COUNT\_MLP}, \textit{BOUNDARY\_MLP}, and \textit{DIST\_MLP} are restricted probes. The remaining models test increasingly rich graph, incidence, cellular, and subcomplex information. The SMCN parameter count depends on the marking configuration. Appendix~\ref{app:controls} reports the corresponding counts.}

\scriptsize
\renewcommand{\arraystretch}{1.12}
\setlength{\tabcolsep}{2pt}

\begin{tabular}{@{}cp{2.45cm}rp{4.00cm}p{3.50cm}@{}}
\toprule
\textbf{Level} &
\textbf{Model} &
\textbf{Params} &
\textbf{Information seen} &
\textbf{Diagnostic role}\\
\midrule

1 &
\textbf{COUNT\_MLP} &
226 &
$(\chi,|V|,|E|,|F|)$ &
C1 count probe\\

2 &
\textbf{BOUNDARY\_MLP} &
226 &
Boundary count and loop lengths &
C4 boundary probe\\

3 &
\textbf{DIST\_MLP} &
482 &
Node-to-cell SPD histogram &
C5 metric probe\\

4 &
\textbf{GIN} &
34{,}574 &
\(1\)-skeleton adjacency &
\(1\)-WL graph baseline\\

5 &
\textbf{ESAN} &
$\approx 70$K &
Ego-net subgraph bag &
Graph substructure baseline\\

6 &
\textbf{HASSE\_GIN} &
34{,}638 &
Ranked Hasse relations &
Typed incidence baseline\\

7 &
\textbf{CIN/HOMP} &
221K/256K &
Unsigned cellular relations &
Cellular message passing\\

8 &
\textbf{SMCN} &
143K--182K &
Face-marked subcomplex lift &
Subcomplex lift model\\

\bottomrule
\end{tabular}
\label{tab:baseline-ladder}
\end{table}

%%%%%%%%%%%%%%%%%%%%%%%%%%%%%%%%%%%%%%%%%%%
\section{Results}
\label{sec:results}
%%%%%%%%%%%%%%%%%%%%%%%%%%%%%%%%%%%%%%%%%%%

\textbf{Experimental Setup.} We use a BREC-style protocol in which \textsc{sep} tests target pair separation at threshold \(\tau=60\), while \textsc{rel} tests stability under input permutations. In the main experiments reported in Table~\ref{tab:brec-results}, all models use hidden dimension \(64\) and train for \(40\) epochs. We evaluate \(64\) cylinder vs. M\"obius, \(64\) torus vs. Klein, and \(50\) punctured pairs, together with \(16\) certified Fundamental Group pairs across four seeds. The baselines include restricted count, boundary, and distance probes, \(1\)-WL-bounded models, and the more expressive ESAN and SMCN. For the Fundamental Group task, SMCN uses either binary face-corner marking or shortest-path marking.

\begin{table*}[t]
\centering
\caption{BREC evaluation results. Orientability entries report distinct benchmark pairs. Fundamental Group entries report pair--seed evaluations over 16 fixed pairs and four seeds; they are not 64 independent pairs. Each solved entry satisfies both target separation and permutation reliability. These values measure pair separability rather than classification accuracy.}
\label{tab:brec-results}
\footnotesize
\renewcommand{\arraystretch}{1.30}
\setlength{\tabcolsep}{5pt}

\resizebox{\linewidth}{!}{%
\begin{tabular}{@{}llcccc@{}}
\toprule

\multirow{3}{*}{\textbf{Level}} &
\multirow{3}{*}{\textbf{Model}} &
\multicolumn{3}{c}{\textbf{Orientability}} &
\multicolumn{1}{c}{\textbf{Fundamental Group}}\\

\cmidrule(lr){3-5}
\cmidrule(lr){6-6}

& &
\shortstack{\textbf{Cylinder vs.}\\\textbf{M\"obius}} &
\shortstack{\textbf{Torus vs.}\\\textbf{Klein}} &
\shortstack{\textbf{Punc.\ Torus vs.}\\\textbf{Punc.\ Klein}} &
\shortstack{\textbf{Surface Group}\\\textbf{vs.\ \(F_4\)}}\\

& &
\textsc{BREC Solved} &
\textsc{BREC Solved} &
\textsc{BREC Solved} &
\textsc{BREC Solved}\\

\midrule

\multirow{3}{*}{
\parbox{1.8cm}{\raggedright\itshape
Probes\\
No message passing}}
&
\texttt{COUNT\_MLP}
& 0/64
& 0/64
& 0/50
& 0/64\\

&
\texttt{BOUNDARY\_MLP}
& \textbf{64/64}
& 0/64
& 0/50
& 0/64\\

&
\texttt{DIST\_MLP}
& \textbf{64/64}
& \textbf{60/64}
& \textbf{45/50}
& \textit{0/64}\\

\addlinespace[2pt]
\midrule

\multirow{3}{*}{
\parbox{1.8cm}{\raggedright\itshape
\(1\) WL bounded}}
&
\texttt{GIN}
& 0/64
& 0/64
& 0/50
& 0/64\\

&
\texttt{HASSE\_GIN}
& 0/64
& 0/64
& 0/50
& 0/64\\

&
\texttt{HOMP/CIN}
& 0/64
& 5/64
& 0/50
& 0/64\\

\addlinespace[2pt]
\midrule

\multirow{3}{*}{
\parbox{1.8cm}{\raggedright\itshape
Beyond \(1\) WL}}
&
\texttt{ESAN}
& 16/64
& 16/64
& 5/50
& \textbf{37/64}\\

&
\texttt{SMCN} (binary marking)
& \textbf{64/64}
& \textbf{64/64}
& 36/50
& 48/64\\

&
\texttt{SMCN} (shortest-path marking)
& \textbf{64/64}
& \textbf{64/64}
& \textbf{49/50}
& \textbf{51/64}\\

\bottomrule
\end{tabular}%
}

\end{table*}

\begin{table*}[t]
\centering
\caption{Diagnostic target and reliability margins for model families.}
\label{tab:brec-margin-summary}
\footnotesize
\renewcommand{\arraystretch}{1.20}
\setlength{\tabcolsep}{5pt}

\resizebox{\linewidth}{!}{%
\begin{tabular}{@{}lllrr@{}}
\toprule
\textbf{Family} &
\textbf{Model class} &
\textbf{Representative} &
\textbf{Target margin} &
\textbf{Reliability margin}\\
\midrule

\multirow{3}{*}{\textbf{FG Genus \(2\)}}
& Probe
& \texttt{COUNT\_MLP}
& \(0.00\pm0.00\)
& \(0.00\pm0.00\)\\

& Local MP
& \texttt{HOMP}
& \(0.08\pm0.01\)
& \(0.07\pm0.01\)\\

& Beyond \(1\) WL
& \textbf{\texttt{SMCN}}
& \(\boldsymbol{1.38\mathrm{e}{3}\pm2.00\mathrm{e}{3}}\)
& \(\boldsymbol{1.09\mathrm{e}{-3}\pm8.92\mathrm{e}{-4}}\)\\

\midrule

\multirow{3}{*}{\textbf{Cyl. vs.\ M\"ob.}}
& Probe
& \textbf{\texttt{DIST\_MLP}}
& \(\boldsymbol{1.66\mathrm{e}{12}\pm1.62\mathrm{e}{2}}\)
& \(\boldsymbol{0.12\pm0.01}\)\\

& Local MP
& \texttt{HOMP}
& \(0.06\pm0.02\)
& \(0.04\pm0.01\)\\

& Beyond \(1\) WL
& \texttt{ESAN}
& \(1.03\mathrm{e}{12}\pm4.23\mathrm{e}{4}\)
& \(7.06\mathrm{e}{-3}\pm1.52\mathrm{e}{-3}\)\\

\midrule

\multirow{3}{*}{\textbf{Punc.\ T vs.\ Punc.\ K}}
& Probe
& \textbf{\texttt{DIST\_MLP}}
& \(\boldsymbol{8.65\mathrm{e}{11}\pm4.04\mathrm{e}{4}}\)
& \(\boldsymbol{0.13\pm0.01}\)\\

& Local MP
& \texttt{HOMP}
& \(0.01\pm0.01\)
& \(0.01\pm0.01\)\\

& Beyond \(1\) WL
& \texttt{SMCN}
& \(6.70\mathrm{e}{4}\pm3.99\mathrm{e}{4}\)
& \(7.88\mathrm{e}{-5}\pm3.44\mathrm{e}{-5}\)\\

\midrule

\multirow{3}{*}{\textbf{Tor. vs.\ Klein}}
& Probe
& \textbf{\texttt{DIST\_MLP}}
& \(\boldsymbol{1.05\mathrm{e}{11}\pm1.08\mathrm{e}{4}}\)
& \(\boldsymbol{0.00\pm0.00}\)\\

& Local MP
& \texttt{HASSE\_GIN}
& \(0.02\pm5.56\mathrm{e}{-4}\)
& \(0.01\pm1.82\mathrm{e}{-3}\)\\

& Beyond \(1\) WL
& \texttt{ESAN}
& \(2.28\mathrm{e}{13}\pm4.16\mathrm{e}{4}\)
& \(0.02\pm0.01\)\\

\bottomrule
\end{tabular}%
}
\end{table*}

\paragraph{Orientability.}
Table~\ref{tab:brec-results} reports the number of pairs satisfying both target separation and permutation reliability. \text{COUNT\_MLP} solves no orientability pairs, whereas \text{BOUNDARY\_MLP} solves all \(64\) cylinder vs. M\"obius pairs. More importantly, \text{DIST\_MLP} solves \(64/64\), \(60/64\), and \(45/50\) pairs in the cylinder vs. M\"obius, torus vs. Klein, and punctured families using only shortest-path histograms. \text{GIN} and \text{HASSE\_GIN} solve no pairs, \text{HOMP/CIN} solves only \(5/64\) torus vs. Klein pairs, and \text{ESAN} separates a subset of the evaluated pairs. \text{SMCN} with binary marking solves \(64/64\), \(64/64\), and \(36/50\) pairs, while shortest-path marking solves \(64/64\), \(64/64\), and \(49/50\). However, as restricted boundary and distance probes have already succeeded, these results cannot be interpreted as true orientability detection.

To examine how separation changes with message-passing depth without supplying explicit shortest-path values, we additionally evaluate binary-marked \texttt{SMCN} at depths \(L\in\{1,2,3,4,5\}\) on the same 50 punctured pairs, using three model seeds per depth. The mean number of solved pairs increases from (0.0) at (L=1) to (25.3) at (L=5), while the largest grid height separated under any seed increases from none to (h=9). This empirical depth-dependent trend is aligned with the audited cycle scale, whose first closed-walk difference occurs at (L=h+1). However, it does not establish that cycles are the causal feature or that a general depth barrier exists.The complete analysis is provided in Appendix~\ref{app:depth-sweep}, with the cycle audit in Table~\ref{tab:cycle-audit}, the per-depth results in Table~\ref{tab:depth-solved}, and the height-stratified results in Table~\ref{tab:depth-height}.

\paragraph{Fundamental Group.}
The benchmark contains $16$ fixed certified pairs, each evaluated under four
model seeds; hence $64$ denotes repeated pair–seed evaluations, not $64$
independent pairs. All three restricted probes and the tested
\textsc{GIN}, \textsc{HASSE\_GIN}, and \textsc{HOMP/CIN} baselines obtain
$0/64$. In contrast, \textsc{ESAN} separates $37/64$ pair–seed evaluations,
while \textsc{SMCN} separates $48/64$ with binary marking and $51/64$ with
shortest-path marking. For shortest-path SMCN, nine of the $16$ unique pairs
are separated by all four seeds, four by three seeds, one by two seeds, one by
one seed, and one by no seed; thus $15/16$ pairs are separated at least once
and $9/16$ consistently across all four seeds. The failure of
\textsc{DIST\_MLP} shows only that its tested aggregated shortest-path
histogram and training budget are insufficient for separation; it does not
establish that the broader metric channel is closed. The binary-marking result
shows that separation is not solely attributable to explicit shortest-path
marking, but it does not exclude other realization-level signals.

These separations occur while integral homology, cell counts, local-incidence
and vertex-link profiles, boundary structure, selected word statistics, and
rank-colored Hasse $1$-WL are matched. However, the ordered signed attaching
word used to certify the mathematical distinction is not supplied to the
models, which instead receive unsigned incidence, adjacency, bridge relations,
and the selected SMCN marking after realization. The models may therefore
exploit another global feature introduced during realization.
Table~\ref{tab:brec-results} reports the primary solved counts, while
Table~\ref{tab:brec-margin-summary} reports diagnostic target and reliability
margins; these margins are not accuracy measures and should not be used alone
to rank models. Overall, the results establish a representation-relative
separation beyond the matched homological, local, and Hasse-$1$-WL controls,
but do not establish that the models compute $\pi_1$ or are directly sensitive to non-Abelian structure.

\section{Conclusion and Future Work}
\label{sec:conclusion}

This paper studied whether topological learning benchmarks measure the invariant they are named after or whether models can solve them through simpler information retained by the representation. Our main contribution is a representation-relative certification framework for synthetic topological benchmarks. The framework identifies shortcut channels, checks them with exact generator-side gates, tests selected channels with restricted probes, and reports pair separation only together with permutation reliability and explicit claim limits.

The first insight is that matching familiar topological quantities is not enough. In the Orientability Family, the cylinder vs. M\"obius, torus vs. Klein, and punctured torus vs. punctured Klein pairs differ in orientability, but they also expose boundary, homology, metric, or cycle information. In particular, a shortest-path probe with no message passing separates most orientability pairs. This shows that positive results on these benchmarks cannot be treated as evidence of orientability detection unless those shortcut channels are removed or controlled.
The second contribution is a Fundamental Group family controlled relative to the listed generator-side channels. We construct genus-\(2\) surface-group versus free-group presentation complexes with different fundamental groups but matched integral homology, cell counts, local incidence profiles, vertex links, boundary structure, word statistics, and rank-colored Hasse \(1\)-WL partitions. On this family, the restricted probes and local-message-passing baselines fail, while richer subgraph and subcomplex models separate a substantial fraction of pairs.
This shows that these architectures can access structure beyond the tested homological and local controls. The impact of the study is methodological. It argues that topological expressivity claims should be made relative to the representation, the controlled channels, and the evaluation protocol. A model separation is scientifically meaningful only after simpler explanations have been tested. This perspective turns benchmark design into an auditable process rather than a naming convention based on the intended invariant.

The work has limitations. The certification is not unconditional: it is relative to the emitted representation, the declared channels, the tested probes, and the generated families. Probe failure does not exclude every possible shortcut, only the tested feature map and model budget. The Fundamental Group result does not show that SMCN computes \(\pi_1\), and the current experiments do not establish generalization to arbitrary presentation complexes or unseen topological families.

Future work should strengthen the generator by adding further metrics, spectral controls, and cycle controls, and should extend SMCN-style architectures with mechanisms that more directly represent signed and ordered attaching information.
A broader benchmark suite over additional presentation complexes would clarify when subcomplex lifting provides genuine sensitivity to non-Abelian structure.

\section*{Author Contributions}

Raghuwansh Raj: Conceptualization, methodology, software, experiments, formal analysis, visualization, and writing of the original draft.
George Panagopoulos: Experimental direction, draft review, and editing.
Jun Pang: Methodological discussions, draft review, and editing.
All authors reviewed and approved the final manuscript.

\section*{Acknowledgments}

The computational experiments were performed on the Luxembourg national supercomputer, HPC MeluXina. The authors gratefully acknowledge the LuxProvide team for their technical support. The work of Georgios Panagopoulos and Jun Pang was supported by the Luxembourg National Research
Fund (FNR) under the grant agreement C25/IS/19480462/GALICIA. %The authors also acknowledge the University of Luxembourg, Department of Computer Science, Laboratory for Cryptography and Security(LACS) for the monetary support.

%%%%%%%%%%%%%%%%%%%%%%%%%%%%%%%%%%%%%%%%%%%

\clearpage

\newpage
\input{appendix}

\end{document}

%% file: appendix.tex
\newcommand{\Cmatch}{$\bullet$}      % matched
\newcommand{\Cpart}{$\circleddash$}  % partial
\newcommand{\Cprobe}{$\circ$}        % probe-tested, not closed
\newcommand{\Cleak}{$\times$}        % leak
\newcommand{\Cuntest}{--}            % untested/open
\appendix
\captionsetup{font=small}

\clearpage
\phantomsection
\pdfbookmark[1]{Appendix Contents}{appendix-contents}
\begin{center}
    \vspace*{0.6in}
    {\Large\bfseries Appendix}\par
    \vspace{0.5em}
    {\large Supplementary Definitions, Constructions, Audits, and Experiments}
\end{center}

\vspace{1.5em}
{\large\bfseries Appendix Contents}\par
\vspace{0.75em}

\begin{tabularx}{\textwidth}{@{}>{\raggedright\arraybackslash}Xr@{}}
\hyperref[app:topological-background]{Appendix~\ref*{app:topological-background}: Mathematical and Representation Background}
& \hyperref[app:topological-background]{\pageref*{app:topological-background}}\\[0.35em]
\hyperref[app:certification]{Appendix~\ref*{app:certification}: Certification Details}
& \hyperref[app:certification]{\pageref*{app:certification}}\\[0.35em]
\hyperref[app:baseline-models]{Appendix~\ref*{app:baseline-models}: Baseline Models and Diagnostic Probes}
& \hyperref[app:baseline-models]{\pageref*{app:baseline-models}}\\[0.35em]
\hyperref[app:orient-construction]{Appendix~\ref*{app:orient-construction}: Orientability Benchmark Construction}
& \hyperref[app:orient-construction]{\pageref*{app:orient-construction}}\\[0.35em]
\hyperref[app:distance-matched-orientability]{Appendix~\ref*{app:distance-matched-orientability}: Distance-Histogram-Matched Orientability Family}
& \hyperref[app:distance-matched-orientability]{\pageref*{app:distance-matched-orientability}}\\[0.35em]
\hyperref[app:depth-sweep]{Appendix~\ref*{app:depth-sweep}: Depth Sweep for the Punctured Orientability Family}
& \hyperref[app:depth-sweep]{\pageref*{app:depth-sweep}}\\[0.35em]
\hyperref[app:fg-construction]{Appendix~\ref*{app:fg-construction}: Fundamental Group Benchmark Construction}
& \hyperref[app:fg-construction]{\pageref*{app:fg-construction}}\\[0.35em]
\hyperref[app:relator]{Appendix~\ref*{app:relator}: Held-Out Relator Family}
& \hyperref[app:relator]{\pageref*{app:relator}}\\[0.35em]
\hyperref[app:experimental-details]{Appendix~\ref*{app:experimental-details}: Experimental and Sanity Check Details}
& \hyperref[app:experimental-details]{\pageref*{app:experimental-details}}\\[0.35em]
\hyperref[app:genai]{GenAI Usage Statement}
& \hyperref[app:genai]{\pageref*{app:genai}}\\
\end{tabularx}

\clearpage

\section{Mathematical and Representation Background}
\label{app:topological-background}

This appendix collects the mathematical definitions, construction details,
certification checks, and auxiliary experiments that support the main text.

\subsection{Finite Cell Complexes and Their Combinatorial Encodings}
\label{app:combinatorial-complexes}

Throughout, \(X\) denotes a finite cell complex together with the combinatorial
data used to encode it. Its cells form a finite graded set
\[
X=\bigsqcup_{k=0}^{d}X_k,
\qquad
X_k=\{\sigma\in X:\rho(\sigma)=k\},
\]
where \(\rho:X\to\{0,\ldots,d\}\) is the rank function, \(X_k\) is the set
of \(k\)-dimensional cells, and \(d=\max_{\sigma\in X}\rho(\sigma)\). For the
cell complexes considered here, rank agrees with cell dimension. We write
\(\sigma\prec\tau\) when \(\sigma\) is a proper face of \(\tau\), that is,
when \(\sigma\subseteq\overline{\tau}\setminus\tau\). In addition to this
incidence relation, the attaching maps are regarded as part of the cell-complex
data; this distinction is important because the face relation alone does not,
in general, determine a non-regular CW complex. We denote the underlying
topological space by \(\operatorname{Real}(X)\).

For \(\sigma\in X\), define its incidence boundary and coboundary sets by
\[
\begin{aligned}
\mathcal B(\sigma)&=\{\tau\in X:\tau\prec\sigma\},\\
\mathcal C(\sigma)&=\{\tau\in X:\sigma\prec\tau\}.
\end{aligned}
\]
These are sets of incident cells, not the point-set boundary of
\(\operatorname{Real}(X)\). More precisely, for \(\sigma\in X_r\), cells of
rank \(r\) are lower adjacent through rank \(s<r\) and upper adjacent through
rank \(t>r\), respectively, when they belong to
\[
\begin{aligned}
\mathcal N^{\downarrow}_{s}(\sigma)
&=\{\tau\in X_r\setminus\{\sigma\}:\\
&\qquad \exists\delta\in X_s,\ \delta\prec\sigma\text{ and }\delta\prec\tau\},\\
\mathcal N^{\uparrow}_{t}(\sigma)
&=\{\tau\in X_r\setminus\{\sigma\}:\\
&\qquad \exists\eta\in X_t,\ \sigma\prec\eta\text{ and }\tau\prec\eta\}.
\end{aligned}
\]
Cellular message passing can use these typed incidence and adjacency relations,
including the usual adjacency of vertices in the \(1\)-skeleton
~\citep{hajij2022tdl,hajij2024topox,bodnar2021cellular}.

The rank-typed incidence graph used in our refinement test is denoted by
\(\mathcal H(X)\) and called the Hasse graph. It is the undirected graph
\[
\begin{aligned}
V(\mathcal H(X))&=X,\\
E(\mathcal H(X))
&=\bigl\{\{\sigma,\tau\}:\sigma\prec\tau,
\ \rho(\tau)=\rho(\sigma)+1\bigr\}.
\end{aligned}
\]
For a graded face poset this is its usual undirected Hasse diagram; the formula
also states explicitly which incidences are retained when the underlying CW
complex is non-regular. In the refinement test, each node is initialized with
its rank \(\rho(\sigma)\). By
\emph{rank-colored one-dimensional Weisfeiler--Leman refinement} we mean
ordinary \(1\)-WL graph refinement on \(\mathcal H(X)\), initialized by these
rank labels.

\subsection{Euler Characteristic, Homology, and Orientability}
\label{app:euler-homology}

The Euler characteristic of the finite cell complex is
\[
\chi\bigl(\operatorname{Real}(X)\bigr)
=\sum_{k=0}^{d}(-1)^k|X_k|.
\]
For brevity, we write this as \(\chi(X)\). For the two-dimensional complexes
used here,
\(\chi(X)=|X_0|-|X_1|+|X_2|\), where \(|X_k|\) is the number of rank-\(k\)
cells.

After choosing an orientation for every cell, the cellular chain group is the
free abelian group \(C_k(X;\mathbb Z)\cong\mathbb Z^{X_k}\). Its cellular
boundary homomorphisms
\[
\partial_k:C_k(X;\mathbb Z)\longrightarrow C_{k-1}(X;\mathbb Z)
\]
are defined by the signed cellular incidence numbers and satisfy
\(\partial_{k-1}\circ\partial_k=0\). We use
\[
H_k(X;\mathbb Z)
:=H_k\bigl(\operatorname{Real}(X);\mathbb Z\bigr)
=\ker(\partial_k)/\operatorname{im}(\partial_{k+1})
\]
for the \(k\)th integral homology group, and
\(\beta_k=\operatorname{rank}_{\mathbb Z}H_k(X;\mathbb Z)\) for its Betti
number. Betti numbers record only the rank of the free part and therefore do
not detect torsion. Every exact homology gate in this study instead computes
the invariant factors of the integer boundary matrices using Smith normal form
~\citep{hatcher2002algebraic,massey1991basic,
kannan1979polynomial,storjohann1996near}.

Let \(M\) be a compact connected surface. A cellulation of \(M\) is
orientable when its \(2\)-cells can be oriented so that the induced
orientations on every interior edge cancel pairwise. If \(\partial M=\varnothing\),
then \(M\) is orientable if and only if
\(H_2(M;\mathbb Z)\cong\mathbb Z\); its generator is the integral fundamental
class. If \(\partial M\neq\varnothing\), then
\(H_2(M;\mathbb Z)=0\) regardless of orientability, whereas \(M\) is orientable
if and only if
\(H_2(M,\partial M;\mathbb Z)\cong\mathbb Z\). Thus absolute integral homology
need not determine orientability for surfaces with boundary
~\citep{hatcher2002algebraic,stillwell1993classical}.

\subsection{Group Presentations and Presentation Complexes}
\label{app:presentation-complexes}

Let
\[
P=\langle g_1,\ldots,g_n\mid r_1,\ldots,r_m\rangle
\]
be a finite group presentation, where each relator \(r_j\) is a word in the
free group \(F_n=F(g_1,\ldots,g_n)\). The presented group is
\[
G(P)=F_n/\langle\!\langle r_1,\ldots,r_m\rangle\!\rangle,
\]
where \(\langle\!\langle\cdot\rangle\!\rangle\) denotes normal closure. Its
presentation complex \(K(P)\) is the two-dimensional CW complex with one
vertex \(v\), one oriented loop for each generator \(g_i\), and one
\(2\)-cell for each relator \(r_j\), whose attaching loop spells \(r_j\). The fundamental
group of the realization, based at \(v\), therefore satisfies
\[
\pi_1\bigl(\operatorname{Real}(K(P)),v\bigr)\cong G(P)
\]
~\citep{hatcher2002algebraic,lyndon1977combinatorial}.

\paragraph{Benchmark topologies at a glance.}
The cylinder and M\"obius band both have homotopy type $S^1$ and integral
homology $(\mathbb Z,\mathbb Z,0)$. The torus and Klein bottle are both closed,
but their integral homology differs. The punctured torus and punctured Klein
bottle both have homotopy type $S^1\vee S^1$ and integral homology
$(\mathbb Z,\mathbb Z^2,0)$, despite differing in orientability. Finally, the
genus-$2$ surface presentation complex and the freely trivial presentation
complex both have integral homology $(\mathbb Z,\mathbb Z^4,\mathbb Z)$, while
their fundamental groups differ. Table~\ref{tab:topology} gives the compact
comparison, and Appendices~\ref{app:orient-construction} and
\ref{app:fg-construction} give the full constructions.

\subsection{From a Complex to the Model Input}
\label{app:machine-representation}

The model does not receive \(\operatorname{Real}(X)\) or the complete
attaching-map data directly. A preprocessing map \(\Phi\) converts \(X\) into
the finite collection of feature and connectivity arrays used by the model;
we denote that collection by \(\Phi(X)\). For ranks
\(r,r_1,r_2\in\{0,\ldots,d\}\) with \(r_1<r_2\), the principal tensors are
summarized in Table~\ref{tab:app-tensors}.

\begin{table*}[!t]
\caption{Connectivity information carried by the emitted machine representation.}
\label{tab:app-tensors}
\centering
\small
\begin{tabular}{@{}ll@{}}
\toprule
\textbf{Tensor} & \textbf{Meaning}\\
\midrule
\texttt{x\_r} & Feature matrix for cells in \(X_r\)\\
\texttt{num\_cells\_r} & Cell count \(|X_r|\)\\
\texttt{edge\_index\_incidence\_r1\_r2} & Pairs \((\sigma,\eta)\in X_{r_1}\times X_{r_2}\) with \(\sigma\prec\eta\)\\
\texttt{edge\_index\_adjacency\_r1\_r2} & Pairs of \(r_1\)-cells sharing an \(r_2\)-cell\\
\texttt{bridge\_index\_adjacency\_r1\_r2} & Mediating \(r_2\)-cell for each preceding adjacency pair\\
\texttt{edge\_index\_coadjacency\_r2\_r1} & Pairs of \(r_2\)-cells sharing an \(r_1\)-cell\\
\texttt{bridge\_index\_coadjacency\_r2\_r1} & Mediating \(r_1\)-cell for each preceding coadjacency pair\\
\bottomrule
\end{tabular}

\end{table*}

The incidence arrays retain only whether an incidence is nonzero; they do not
retain the signed integer coefficients of the cellular boundary operators.
Whenever a face is supplied by its vertices, those vertices are represented as
an unordered set. Consequently, orientation signs, cyclic orders, traversal
multiplicities, and the original ordered attaching words are not supplied
\emph{directly} to the tested models, although consequences of that data may
still be recoverable indirectly from the remaining unsigned relations. If a
model separates \(\Phi(X)\) from \(\Phi(Y)\), this establishes sensitivity to
the chosen encodings for that pair. It does not, by itself, establish that the
model computes a topological invariant \(\tau\) on a broader class of
realizations.

\section{Certification Details}
\label{app:certification}

\subsection{Representation-Relative Certification}
\label{app:representation-certification}

Let \(\tau\) be the target invariant and let
\(\mathcal C=\{C_1,\ldots,C_m\}\) be the audited shortcut channels. A
benchmark pair \((X,Y)\) must satisfy
\[
\tau(\operatorname{Real}(X))
\neq
\tau(\operatorname{Real}(Y)).
\]
For every controlled channel \(C_j\), the corresponding information must be
matched,
\[
C_j(\Phi(X))=C_j(\Phi(Y)),
\]
or its status must be reported as leaking, partial, or open. Certification is
therefore relative to \((\Phi,\mathcal C)\); it is not an unconditional
statement that every possible shortcut has been removed.

\subsection{Channel Taxonomy}
\label{app:shortcut-channels}

Let
\[
\mathcal{C}=\{C_1,\ldots,C_8\}
\]
denote the shortcut channels considered in our benchmark audit. Each channel
represents information that may allow a model to distinguish a pair
\((X,Y)\) without computing the intended target invariant. Since the model
receives the machine representation \(\Phi(X)\), all channel assessments are
relative to this representation and to the explicitly declared tests.

\begin{table*}[t]
\caption{
Shortcut channels considered in the representation-relative benchmark audit.
Each channel provides a possible alternative explanation for model separation.
}
\label{tab:shortcut-channels}
\centering
\footnotesize
\setlength{\tabcolsep}{4pt}
\renewcommand{\arraystretch}{1.15}

\begin{tabularx}{\linewidth}{
    @{}
    c
    >{\raggedright\arraybackslash}p{0.18\linewidth}
    >{\raggedright\arraybackslash}X
    @{}
}
\toprule
\textbf{Channel} & \textbf{Name} & \textbf{Description}\\
\midrule

\(C_1\)
&
Cell counts
&
\textbf{Checks:} Numbers of vertices, edges, and faces, together with
\(\chi(X)=|X_0|-|X_1|+|X_2|\).
\newline
\textbf{Why it matters:} Different cell counts or Euler characteristics may
allow separation using size information alone.
\\

\(C_2\)
&
Homology
&
\textbf{Checks:} Integral homology groups \(H_k(X;\mathbb Z)\), including
Betti numbers and torsion.
\newline
\textbf{Why it matters:} If homology differs, the task may be solved using an
abelian invariant rather than the intended harder invariant.
\\

\(C_3\)
&
Local incidence
&
\textbf{Checks:} Vertex degrees, edge--face incidence counts, cell-incidence
profiles, and vertex-link profiles.
\newline
\textbf{Why it matters:} Local message-passing models may exploit these
neighborhood statistics without detecting the target global invariant.
\\

\(C_4\)
&
Boundary
&
\textbf{Checks:} Numbers of boundary components, boundary-loop lengths, and
related boundary statistics.
\newline
\textbf{Why it matters:} Boundary differences can provide simple shortcuts,
especially for surfaces with boundary.
\\

\(C_5\)
&
Global metric
&
\textbf{Checks:} Shortest-path distances, node-to-cell distance histograms,
eccentricities, and related metric summaries.
\newline
\textbf{Why it matters:} A model may distinguish different discretization
geometries without detecting the intended topological invariant.
\\

\(C_6\)
&
Spectral
&
\textbf{Checks:} Spectra of selected adjacency, Hasse-graph, or Laplacian
operators.
\newline
\textbf{Why it matters:} Spectral information may encode global structure
that is not visible in cell counts or local-incidence statistics.
\\

\(C_7\)
&
Short cycles
&
\textbf{Checks:} Girth, short-cycle counts, cycle-length distributions, and
cycles visible within the tested message-passing depth.
\newline
\textbf{Why it matters:} Finite-depth message passing may detect short cycles,
so unequal cycle statistics can explain model separation.
\\

\(C_8\)
&
Rank-colored Hasse \(1\)-WL
&
\textbf{Checks:} One-dimensional Weisfeiler--Leman refinement on the Hasse
graph \(\mathcal H(X)\), initialized with cell-rank labels.
\newline
\textbf{Why it matters:} This determines whether typed incidence refinement
already distinguishes the pair before richer models are applied.
\\

\bottomrule
\end{tabularx}

\end{table*}
\paragraph{Interpretation of channel status.}
For each benchmark pair, every channel is assigned one of the statuses used
below. A channel is labeled
\textit{``matched''} when all declared statistics for that channel agree between
the two members of the pair. It is labeled a \textit{leak} when a measured
difference or a successful restricted probe provides an alternative
explanation for separation. The label \textit{partial} indicates that only
some aspects of the channel have been tested, whereas \textit{open} indicates
that the channel remains insufficiently examined. At the aggregate-channel
level, a partial status may contain a mixture of matched and leaking audited
subchannels.

Throughout the tables, we use
\Cmatch{} for \emph{matched},
\Cleak{} for a \emph{leak},
\Cpart{} for \emph{partial},
\Cprobe{} for \emph{probe-tested but not closed}, and
\Cuntest{} for \emph{open or untested}.
Probe failure is an experimental outcome rather than evidence that a channel
is matched.
\subsection{Audited Shortcut Channels}
\label{app:channels}

An exact gate is evaluated inside the generator, and a candidate pair is
emitted only if the gate passes. A restricted probe is a trainable diagnostic
that receives only one named channel. A successful probe supplies a sufficient
alternative explanation for model separation. A failed probe has a narrower
meaning: it rules out only the tested feature map, architecture, optimization
budget, and parameter range.
\paragraph{Shortcut-channel audit.}
A difference in the target invariant is not sufficient to establish that a
model has detected that invariant. The model may instead exploit a simpler
difference in the representation, such as cell counts, boundary structure,
local incidence patterns, or graph distances. We therefore audit the eight
shortcut channels \(C_1,\ldots,C_8\) listed in
Table~\ref{tab:shortcut-channels}. Channels \(C_1\)--\(C_4\) cover cell counts,
homology, local incidence, and boundary information. Channels \(C_5\)--\(C_7\)
cover global metric, spectral, and short-cycle information. Channel \(C_8\)
tests whether rank-colored \(1\)-WL refinement on the Hasse graph already
distinguishes the pair. Together, these channels provide a structured account
of alternative explanations for model separation. 
\paragraph{Offline gate} An offline gate is a deterministic check performed during dataset generation. A candidate pair is accepted only when every required gate passes. These checks enforce properties such as equal cell counts, exact integral homology, matching incidence and boundary profiles, word statistics, and rank-colored Hasse \(1\)-WL equivalence. Passing a gate proves equality only for the statistic it tests; it does not guarantee that the representation contains no finer shortcut.

\paragraph{Empirical probe} An empirical probe is a restricted trainable model that receives information from only one specified channel. For example, \textbf{COUNT\_MLP} receives cell counts, \textbf{BOUNDARY\_MLP} receives boundary statistics, and \textbf{DIST\_MLP} receives shortest-path histograms. If a probe reliably separates a pair, that channel provides a sufficient alternative explanation for the success of a larger model. If it fails, the conclusion is narrower: only the tested feature map, model class, and training budget have been ruled out.

\subsection{Common Model Inputs}
\label{app:common-model-inputs}

Every model is evaluated on the same underlying combinatorial complex and pair ordering. Cells of the same rank receive identical initial features, so the models cannot use identifiers, coordinates, labels, or hand-designed topological features. The common representation contains cell counts, unsigned incidence relations, derived adjacency and coadjacency relations, and bridge indices identifying the cells that mediate these relations. Orientation signs, cyclic orderings, and the original attaching words are not supplied directly.

The models differ only in how they access this common structure. GIN receives the \(1\)-skeleton, Hasse GIN receives the rank-colored Hasse graph, and HOMP/CIN receives the unsigned higher-order cellular relations. ESAN processes a collection of graph substructures, while SMCN constructs face-indexed subcomplex copies. In SMCN, binary marking indicates whether a vertex is incident to the selected face, whereas shortest-path marking records its distance from that face. These markings are reported separately because they provide different information.

Each complex is also expanded into independently permuted copies, with every incidence, adjacency, and bridge tensor relabeled consistently. The target comparison uses copies of the two different complexes, while the reliability comparison uses permuted copies of the same complex. This ensures that a pair is counted as solved only when its separation reflects structural information rather than sensitivity to the ordering of cells.

\subsection{Pair Separation and Reliability}
\label{app:brec-protocol}

Models are
evaluated with the pairwise protocol adapted from BREC~\citep{wang2024brec}.
For each pair, \(S_{\mathrm{target}}\) compares embeddings of the two target
complexes, while \(S_{\mathrm{rel}}\) compares independently permuted copies
of the same complex. At threshold \(\tau_{\mathrm{BREC}}=60\), a pair is
counted as solved only if
\[
S_{\mathrm{target}}>\tau_{\mathrm{BREC}}
\qquad\text{and}\qquad
S_{\mathrm{rel}}\leq\tau_{\mathrm{BREC}}.
\]
The statistics are based on Hotelling \(T^2\) and Mahalanobis distance
~\citep{hotelling1931generalization,mahalanobis1936generalized}. This protocol
measures pair separability, not classification accuracy or generalization.

\section{Baseline Models and Diagnostic Probes}
\label{app:baseline-models}

The models are grouped according to the information they receive rather than
their parameter counts. The first group contains restricted probes with no
message passing. The second group contains models bounded by graph or cellular
refinement procedures. The final group contains architectures that use
subgraph or subcomplex lifts and can exceed ordinary \(1\)-WL expressivity.

\subsection{Restricted Diagnostic Probes}
\label{app:probe-models}

A diagnostic probe is a small multilayer perceptron (MLP) trained on the same
pair-separation task as the main architectures but given access to only one
predefined information channel. These probes are introduced in this study and
do not require external architectural citations.

\textbf{COUNT\_MLP} receives only
\[
\bigl(\chi(X),|X_0|,|X_1|,|X_2|\bigr),
\]
where \(\chi(X)\) is the Euler characteristic and \(|X_k|\) is the number of
rank \(k\) cells. It contains no incidence, adjacency, or metric information.
Its purpose is to determine whether a benchmark pair can be separated using
only basic size statistics. If COUNT\_MLP succeeds, the result can be
explained by the cell-count channel \(C_1\).

\textbf{BOUNDARY\_MLP} receives only the number of boundary components and
their lengths. It does not receive the interior incidence structure of the
complex. The model targets the boundary channel \(C_4\). Its success on the
cylinder versus M\"obius family shows that the different boundary structures
are sufficient for separation. Its zero separation on two closed
surfaces also provides a negative control because both inputs then have an
identically zero boundary feature vector.

\textbf{DIST\_MLP} receives a histogram of shortest-path distances between
nodes and cells in the emitted representation. It performs no message passing
and receives no orientation or attaching-word information. The probe targets
the global metric channel \(C_5\). Its design is motivated by the use of
shortest-path distances as global positional information in graph
architectures~\citep{ying2021graphormer}. If DIST\_MLP separates a pair, shortest-path information alone is a
sufficient explanation for that separation. If it fails, the conclusion is
limited to the tested distance histogram, MLP architecture, and training
budget; it does not prove that every metric statistic is matched.

\subsection{Graph and Cellular Refinement Baselines}
\label{app:local-baselines}

These models perform local message passing. Their receptive fields grow with
depth, and their separating power is related to Weisfeiler--Leman refinement
on the representation they receive.

\paragraph{GIN}
The Graph Isomorphism Network (GIN) operates on the \(1\)-skeleton of the
complex~\citep{xu2019powerful}. It updates each vertex representation using
an injective aggregation of neighboring vertex features:
\[
h_v^{(\ell+1)}
=
\operatorname{MLP}^{(\ell)}
\left(
(1+\epsilon^{(\ell)})h_v^{(\ell)}
+
\sum_{u\in\mathcal N(v)}h_u^{(\ell)}
\right).
\]
Here, \(h_v^{(\ell)}\) is the representation of vertex \(v\) at layer
\(\ell\), and \(\mathcal N(v)\) is its graph neighborhood. Under standard
assumptions, GIN matches the distinguishing power of one-dimensional
Weisfeiler--Leman refinement (\(1\)-WL). It tests whether the ordinary graph
structure of the \(1\)-skeleton is sufficient.

\paragraph{Hasse GIN}
\textbf{HASSE\_GIN} first converts the complex into its rank-colored Hasse
graph. Every vertex, edge, and face of the original complex becomes a node,
and an edge is added between immediately incident cells. GIN message passing
is then applied to this graph, with the cell rank used as the initial node
type.

This baseline sees higher-dimensional cells but remains bounded by \(1\)-WL
on the ranked Hasse representation. It tests whether exposing all cells as
graph nodes is sufficient, without introducing a more expressive cellular or
subcomplex computation. Hasse-graph representations of cell complexes are
also used in topological learning software~\citep{hajij2024topox}; the GIN
expressivity statement follows from~\citep{xu2019powerful}.

\paragraph{CIN}
The Cellular Isomorphism Network (CIN) updates representations of cells using
their cellular neighborhoods~\citep{bodnar2021cellular}. For a cell
\(\sigma\), these may include its boundary, coboundary, lower adjacent, and
upper adjacent cells. A schematic update is
\[
h_\sigma^{(\ell+1)}
=
\operatorname{UPD}^{(\ell)}
\left(
h_\sigma^{(\ell)},
\operatorname{AGG}_{\tau\in\mathcal N(\sigma)}
M^{(\ell)}(h_\sigma^{(\ell)},h_\tau^{(\ell)})
\right),
\]
where \(\mathcal N(\sigma)\) contains the selected cellular neighbors.

CIN is more informative than message passing on the \(1\)-skeleton because it
directly processes higher-rank cells. Its expressivity is related to cellular
Weisfeiler--Leman refinement. In our experiments, it receives unsigned cellular
relations and therefore does not directly receive signed boundary operators.

\paragraph{HOMP}
Higher-Order Message Passing (HOMP) is used as a general local cellular
baseline~\citep{eitan2024blindspots}. It maintains separate representations
for cells of different ranks and exchanges messages through incidence,
adjacency, and coadjacency relations. Unlike Hasse GIN, these relations are
processed according to their cellular type rather than being collapsed into
one graph.

HOMP tests whether unsigned local interactions between higher-dimensional
cells are sufficient for separation. Increasing its depth enlarges its
receptive field, but does not necessarily overcome a refinement bound once
the corresponding rank-colored partition has stabilized. Its failure rules
out only the tested local higher-order mechanism, depth, and optimization
budget.

\subsection{Architectures Beyond Ordinary \texorpdfstring{$\boldsymbol{1}$}{1}-WL}
\label{app:beyond-wl-models}

These architectures construct collections of subgraphs or subcomplexes before
message passing. This allows them to retain relational information that may
be unavailable to ordinary \(1\)-WL message passing. We do not claim that
either architecture is exactly equivalent to a fixed \(k\)-WL test.

\paragraph{ESAN}
The Equivariant Subgraph Aggregation Network (ESAN) represents a graph through
a collection of selected subgraphs~\citep{bevilacqua2022esan}. Each subgraph
is processed by a shared graph network, and the resulting representations are
aggregated into an invariant graph representation. In our experiments, ESAN
uses ego-net subgraphs and the Deep Sets for Subgraphs (DSS) aggregation
scheme.

By preserving which nodes and edges occur together within each selected
subgraph, ESAN can distinguish some graphs that ordinary \(1\)-WL cannot.
However, it operates on graph substructures derived from the \(1\)-skeleton
and does not directly process the complete cellular incidence structure.

\paragraph{SMCN}
The Scalable Multi-Cellular Network (SMCN) constructs a collection of
face-indexed subcomplex representations~\citep{eitan2024blindspots}. For each
selected face \(f\), the model creates a marked copy of the vertex-level
structure. Message passing is performed within each copy, while additional
fiber relations exchange information between copies associated with the same
underlying cell.

The initial marking is implemented in two ways. Binary marking records only
whether a vertex is incident to the selected face:
\[
m_{\mathrm{bin}}(v,f)
=
\mathbf 1\{v\prec f\}.
\]
Shortest-path marking records the distance from a vertex to the selected face:
\[
m_{\mathrm{dist}}(v,f)
=
\min_{u\prec f}d(v,u),
\]
where \(d(v,u)\) is the shortest-path distance in the chosen graph
representation.

Binary marking supplies local incidence information, so global distinctions
must be constructed through message passing. Shortest-path marking provides a
global positional encoding at the input layer. The two settings are evaluated
separately because they expose different information channels.

SMCN is the tested architecture with the richest explicitly supplied
subcomplex structure because it preserves the identity of the marked subcomplex
while performing message passing. Its
success on pairs for which GIN, Hasse GIN, and HOMP fail demonstrates a realized
expressivity separation under the tested representation and training
conditions. It does not prove that SMCN computes the target topological
invariant.

\subsection{Interpretation of the Baseline Ladder}
\label{app:baseline-interpretation}

The baseline ladder supports a one-way inference. If a restricted probe
separates a pair, the same channel provides a sufficient alternative
explanation for any larger model that can access it. For example, if
\textbf{DIST\_MLP} separates a pair and SMCN receives shortest-path marking,
SMCN's success cannot be attributed specifically to the target invariant.

The converse does not hold. Failure of a probe or local baseline does not prove
that a more expressive model computes the invariant. It shows only that the
tested lower level explanation was insufficient. Optimization, depth, feature
design, and untested representation channels remain possible explanations.

% ============================================================
\section{Orientability Benchmark Construction}
\label{app:orient-construction}
% ============================================================

This section describes the common grid construction underlying the cylinder,
M\"obius band, torus, Klein bottle, and punctured surface pairs. The members
differ only in their boundary closure and in whether the seam identification
preserves or reverses orientation.

\subsection{The Rectangular Cell Grid}
\label{app:cell-grid}
Each grid is written as \([h]\times[p]\), where \([h]=\{1,\ldots,h\}\), \([p]=\{1,\ldots,p\}\), \(h\) is the grid height, and \(p\) is the seam length. One member uses an orientation-preserving identification, while the other uses an orientation-reversing identification. The nuisance channels \(C_1\) through \(C_8\) are those defined in the preceding channel taxonomy. For homology, we write \(H_*=(H_0,H_1,H_2)\), where \(H_i\) denotes the \(i\)th integral homology group.
Fix integers \(h,p\geq 4\). The parameter \(h\) is the grid height, and \(p\)
is the seam length. Before the seam is identified, consider the rectangular
square complex with vertices
\[
v_{i,j},
\qquad
0\leq i<h,
\qquad
0\leq j\leq p.
\]
Its \(1\) cells join horizontally or vertically adjacent vertices, and its
\(2\) cells are the unit squares
\[
[v_{i,j},v_{i,j+1},v_{i+1,j+1},v_{i+1,j}],
\qquad
0\leq i<h-1,\quad 0\leq j<p.
\]

The two vertical sides are identified through one of the maps
\[
\rho_0(i)=i,
\qquad
\rho_1(i)=h-1-i.
\]
More explicitly,
\[
v_{i,p}\sim v_{\rho_\varepsilon(i),0},
\qquad
\varepsilon\in\{0,1\}.
\]
The value \(\varepsilon=0\) gives the untwisted identification, while
\(\varepsilon=1\) gives the twisted identification:
\[
\begin{aligned}
\text{untwisted:}\quad
&v_{i,p}\sim v_{i,0},\\
\text{twisted:}\quad
&v_{i,p}\sim v_{h-1-i,0}.
\end{aligned}
\]

The untwisted map preserves the order of the rows across the seam. The twisted
map reverses this order. A path crossing the untwisted seam returns with its
transverse orientation unchanged, whereas crossing the twisted seam reverses
that orientation. This difference produces the orientable and nonorientable
members of each pair.

After the seam identification, the vertex set can be written as
\[
X_0=\{v_{i,j}:0\leq i<h,\ 0\leq j<p\},
\qquad |X_0|=hp.
\]
There are \(hp\) edges in the periodic seam direction. If the row direction
is left open, there are \((h-1)p\) additional row edges. If the row
direction is closed, another \(p\) wrap edges are added.

\subsection{Members with Boundary}
\label{app:bounded-orientability}

Leaving the row direction open gives the bounded pair. We denote the
untwisted member by \(B^0_{h,p}\) and the twisted member by \(B^1_{h,p}\):
\[
B^0_{h,p}\cong S^1\times[0,1],
\qquad
B^1_{h,p}\cong\text{M\"obius band}.
\]

Thus, \(B^0_{h,p}\) is a cylinder and \(B^1_{h,p}\) is a M\"obius band. Their
cell counts are
\[
|X_0|=hp,
\qquad
|X_1|=hp+(h-1)p=2hp-p,
\qquad
|X_2|=(h-1)p.
\]
Consequently,
\[
\chi(X)=|X_0|-|X_1|+|X_2|=0
\]
for both members.

Both spaces deformation retract onto their central circle. Therefore,
\[
B^0_{h,p}\simeq S^1,
\qquad
B^1_{h,p}\simeq S^1,
\]
and
\[
\pi_1(B^0_{h,p})
\cong
\pi_1(B^1_{h,p})
\cong
\mathbb Z.
\]
Their integral homology groups are also identical:
\[
H_*(B^0_{h,p};\mathbb Z)
\cong
H_*(B^1_{h,p};\mathbb Z)
\cong
(\mathbb Z,\mathbb Z,0).
\]

The boundary structures are different. In the cylinder, the top and bottom
rows remain separate and form two boundary loops of length \(p\). Under the
twisted identification, these two rows join into one boundary loop of length
\(2p\). Hence,
\[
\partial B^0_{h,p}\cong S^1\sqcup S^1,
\qquad
\partial B^1_{h,p}\cong S^1.
\]
This difference explains why the boundary probe separates the entire
cylinder versus M\"obius family. It is an intrinsic topological difference
between these surfaces and cannot be removed merely by changing the
triangulation.

\subsection{Closed Members}
\label{app:closed-orientability}

The closed pair is obtained by adding row wrap edges and faces. Equivalently,
the row index is interpreted modulo \(h\). We write
\[
C^0_{h,p}\cong T^2,
\qquad
C^1_{h,p}\cong K,
\]
where \(T^2\) is the torus and \(K\) is the Klein bottle.

Both members have
\[
|X_0|=hp,
\qquad
|X_1|=2hp,
\qquad
|X_2|=hp,
\]
and therefore
\[
\chi(C^0_{h,p})
=
\chi(C^1_{h,p})
=
0.
\]
Both are closed, so
\[
\partial C^0_{h,p}
=
\partial C^1_{h,p}
=
\varnothing.
\]
Their boundary feature vectors are therefore identically zero.

The torus is orientable and has
\[
\pi_1(T^2)
=
\langle a,b\mid aba^{-1}b^{-1}\rangle
\cong
\mathbb Z^2,
\]
with integral homology
\[
H_*(T^2;\mathbb Z)
=
(\mathbb Z,\mathbb Z^2,\mathbb Z).
\]

The Klein bottle is nonorientable and has
\[
\pi_1(K)
=
\langle a,b\mid aba^{-1}=b^{-1}\rangle.
\]
Its integral homology is
\[
H_*(K;\mathbb Z)
=
(\mathbb Z,\mathbb Z\oplus\mathbb Z_2,0).
\]
Therefore, the torus and Klein bottle are not homology matched. They differ
both in the torsion subgroup of \(H_1\) and in \(H_2\). Recovering
this homological distinction from cellular boundary data requires the signed
incidences omitted from the model input
~\citep{hatcher2002algebraic,stillwell1993classical}.

\subsection{Punctured Members}
\label{app:punctured-orientability}

The punctured pair is obtained by removing the interior of one quadrilateral
\(2\) cell from each closed member while keeping its four boundary edges. We
write
\[
P^0_{h,p}
=
C^0_{h,p}\setminus\mathring{e}^2,
\qquad
P^1_{h,p}
=
C^1_{h,p}\setminus\mathring{e}^2.
\]
These are respectively a punctured torus and a punctured Klein bottle.

Removing one face changes only the number of \(2\) cells:
\[
|X_0|=hp,
\qquad
|X_1|=2hp,
\qquad
|X_2|=hp-1.
\]
Hence,
\[
\chi(P^0_{h,p})
=
\chi(P^1_{h,p})
=
-1.
\]
Both members have one boundary component, represented by the four edges of
the removed face.

Both punctured surfaces deformation retract onto a wedge of two circles:
\[
P^0_{h,p}
\simeq
P^1_{h,p}
\simeq
S^1\vee S^1.
\]
It follows that
\[
\pi_1(P^0_{h,p})
\cong
\pi_1(P^1_{h,p})
\cong
F_2,
\]
where \(F_2\) is the free group on two generators, and
\[
H_*(P^0_{h,p};\mathbb Z)
\cong
H_*(P^1_{h,p};\mathbb Z)
\cong
(\mathbb Z,\mathbb Z^2,0).
\]

No homotopy invariant can distinguish this pair because the two members have
the same homotopy type. Nevertheless, the punctured torus is orientable and
the punctured Klein bottle is nonorientable. This makes the punctured pair the
strongest orientability construction in the family. Its remaining metric and
depth-dependent cycle differences must therefore be treated as genuine
alternative explanations.
\begin{table*}[!t]
\caption{Properties of the orientability constructions. The cell-count tuple
lists the numbers of vertices, edges, and faces, respectively.}
\label{tab:app-orientability-members}
\centering
\scriptsize
\setlength{\tabcolsep}{4pt}
\renewcommand{\arraystretch}{1.15}

\begin{tabular}{@{}lcccll@{}}
\toprule
\textbf{Complex}
& \(\boldsymbol{(|X_0|,|X_1|,|X_2|)}\)
& \(\boldsymbol{\chi}\)
& \(\boldsymbol{H_*=(H_0,H_1,H_2)}\)
& \(\boldsymbol{\pi_1}\)
& \textbf{Boundary}\\
\midrule

Cylinder
& \((hp,\,2hp-p,\,(h-1)p)\)
& \(0\)
& \((\mathbb Z,\mathbb Z,0)\)
& \(\mathbb Z\)
& \(2\) loops, length \(p\)\\

M\"obius band
& \((hp,\,2hp-p,\,(h-1)p)\)
& \(0\)
& \((\mathbb Z,\mathbb Z,0)\)
& \(\mathbb Z\)
& \(1\) loop, length \(2p\)\\

\addlinespace[2pt]

Torus
& \((hp,\,2hp,\,hp)\)
& \(0\)
& \((\mathbb Z,\mathbb Z^2,\mathbb Z)\)
& \(\mathbb Z^2\)
& None\\

Klein bottle
& \((hp,\,2hp,\,hp)\)
& \(0\)
& \((\mathbb Z,\mathbb Z\oplus\mathbb Z_2,0)\)
& \(\langle a,b\mid aba^{-1}=b^{-1}\rangle\)
& None\\

\addlinespace[2pt]

Punctured torus
& \((hp,\,2hp,\,hp-1)\)
& \(-1\)
& \((\mathbb Z,\mathbb Z^2,0)\)
& \(F_2\)
& \(1\) loop, length \(4\)\\

Punctured Klein bottle
& \((hp,\,2hp,\,hp-1)\)
& \(-1\)
& \((\mathbb Z,\mathbb Z^2,0)\)
& \(F_2\)
& \(1\) loop, length \(4\)\\

\bottomrule
\end{tabular}

\end{table*}
\subsection{Metric and Cycle Effects}
\label{app:orientability-shortcuts}

Although the paired complexes agree on several topological and combinatorial
statistics, the twisted gluing changes paths crossing the seam. Under the
untwisted map, a path returns to the same row, whereas under the twisted map
it returns to the reflected row. Consequently, shortest routes between cells
can have different lengths.

These graph distances depend on the chosen discretization and are not
topological invariants of the underlying surface. Nevertheless, they are
present in \(\Phi(X)\) and can be extracted from its incidence and adjacency
relations. This explains why a distance-only probe can distinguish most
orientability pairs without receiving orientation signs.

The twisted construction can also alter the lengths of cycles crossing the
seam. The generator matches the shortest prescribed vertical cycle, but the
audit identifies a remaining depth-dependent cycle signal. Therefore, the
cycle channel is only partially controlled. These findings do not invalidate
the mathematical constructions; they limit the invariant-level interpretation
of model separation.

The evaluated grids contain \(64\) cylinder versus M\"obius pairs,
\(64\) torus versus Klein pairs, and \(50\) punctured pairs. Their full
pair-separation results are reported in Table~\ref{tab:brec-results}.
\subsection{Interpretation of the Three Pairs}
\label{app:orient-audit}

\textbf{Cylinder versus M\"obius band.}
Both spaces retract to \(S^1\), and their cell counts, homology, and local
incidence profiles agree. Their boundaries do not agree: the cylinder has two
loops of length \(p\), whereas the M\"obius band has one loop of length
\(2p\). The boundary and global metric channels therefore provide alternative
explanations for separation.

\textbf{Torus versus Klein bottle.}
Both spaces are closed and have the same cell counts and Euler characteristic.
Their integral homology differs, and the tested distance histograms also
separate most pairs. This family is therefore neither homology matched nor
metric controlled.

\textbf{Punctured torus versus punctured Klein bottle.}
Both spaces have one boundary component, homology
\((\mathbb Z,\mathbb Z^2,0)\), and homotopy type \(S^1\vee S^1\).
Counts, local incidence, and boundary statistics are matched. Nevertheless,
the tested distance histograms separate most pairs, and a depth-dependent
cycle signal leaves the cycle channel only partially controlled.

The reported sweeps use \(h,p\in\{4,\ldots,11\}\) for the \(64\) bounded and
closed pairs. The \(50\) punctured pairs use
\(h\in\{4,\ldots,13\}\) and \(p\in\{4,6,8,10,12\}\). The full separation
results are reported in Table~\ref{tab:brec-results}.

\section{A Distance-Histogram-Matched Orientability Family}
\label{app:distance-matched-orientability}

The original punctured-torus versus punctured-Klein-bottle benchmark controls
homology, homotopy type, cell counts, vertex degrees, and boundary statistics.
Nevertheless, a distance-histogram probe separates 45 of the 50 original
pairs.  Thus, although the two classes differ in the intended topological
property, the original discretization admits a simpler metric shortcut.  This
section describes a revised family in which the input to that probe is equal
by construction.  We use the term \emph{distance-histogram matched} rather
than \emph{isometric}: the construction nullifies the stated probe but does
not assert equality of the complete metric or cycle structure.

\subsection{Cell-Complex Construction}

Let $h \geq 6$ be even and set $p=4$.  Begin with an $h\times p$
quadrangulated strip with vertex set
\begin{equation}
    V_{h,p}=\{0,\ldots,h-1\}\times\mathbb{Z}_{p}.
\end{equation}
Horizontal neighbors are joined cyclically in $\mathbb{Z}_{p}$, and adjacent
rows are joined by vertical edges. The row direction is then closed by adding
seam edges and quadrilateral faces from the last row to the first according to
one of two maps on the column coordinate:
\begin{align}
    \sigma_{\mathrm{T}}(j) &= j+2 \pmod 4,
    &\text{(orientation preserving)},\\
    \sigma_{\mathrm{K}}(j) &= 3-j \pmod 4,
    &\text{(orientation reversing)}.
\end{align}
The mapping torus of $\sigma_{\mathrm{T}}$ is a torus, since a half rotation
of the circle is orientation preserving.  The mapping torus of
$\sigma_{\mathrm{K}}$ is a Klein bottle, since the gluing is a reflection.
In each complex, we remove the quadrilateral selected by the same fixed
row-and-column indexing rule in the generator, away from the seam. This
produces one boundary component of length four without deleting any vertices
or edges.

For both classes the cell counts are
\begin{equation}
    (n_0,n_1,n_2)=(4h,8h,4h-1),
\end{equation}
and hence
\begin{equation}
    \chi = n_0-n_1+n_2=-1.
\end{equation}
Both punctured surfaces deformation retract onto a wedge of two circles, so
their integral homology groups agree.  Their vertex-degree profiles also
agree, and their boundary signatures consist of one four-edge loop.  The
restriction $h\geq6$ keeps the removed face disjoint from the seam.  Requiring
$h$ to be even eliminates the parity-dependent distance discrepancy observed
for odd heights.

\subsection{The Controlled Distance Statistic}

Let $G_X$ denote the rank-zero adjacency graph of a cell complex $X$.  For a
vertex $u$ and a two-cell $f$, define the vertex-to-cell distance by
\begin{equation}
    d_X(u,f)=\min_{v\in V(f)}d_{G_X}(u,v),
\end{equation}
where $V(f)$ is the set of vertices incident to $f$.  The untruncated
distance histogram is
\begin{equation}
    H_X(r)=
    \left|\left\{(u,f)\in X_0\times X_2:
    d_X(u,f)=r\right\}\right|,
    \qquad r\in\mathbb{Z}_{\geq0}.
    \label{eq:node-face-histogram}
\end{equation}
This statistic deliberately discards the identities of vertices and faces;
it retains only the multiplicity of each distance.

For the two complexes above, direct counting of distance layers gives, with
$F=4h-1$,
\begin{equation}
H_X(r)=
\begin{cases}
    4F, & r=0,\\
    8F, & 1\leq r<h/2,\\
    4F, & r=h/2,\\
    0,  & r>h/2.
\end{cases}
\label{eq:matched-histogram-closed-form}
\end{equation}
The expression is the same for the punctured torus and punctured Klein
bottle.  Intuitively, when $p=4$, both seam maps are fixed-point-free
involutions on the columns.  For even $h$, their row-distance layers have the
same cardinalities after aggregation over all vertices and remaining 
faces.  The maps differ in orientation, but orientation is not retained by
the aggregate statistic in Eq.~\eqref{eq:node-face-histogram}.

As a concrete example, for $(h,p)=(6,4)$, both classes produce
\begin{equation}
    H_X=(92,184,184,92,0,\ldots),
\end{equation}
where the entries sum to
$24\cdot23=552=|X_0||X_2|$.  In contrast, the original $(6,6)$ pair gives
\begin{align}
    H_{\mathrm{T}}&=(140,280,420,420,0,\ldots),\\
    H_{\mathrm{K}}&=(140,280,420,326,94,0,\ldots),
\end{align}
which exposes a direct classification cue.

The implemented distance probe uses a capped version of this histogram with
the maximum recorded distance $B=10$.  Since Eq.~\eqref{eq:matched-histogram-closed-form}
holds before truncation, any deterministic clipping or tail aggregation
preserves equality.  Consequently, for every generated target pair,
\begin{equation}
    \Phi_{\mathrm{dist}}(X_{\mathrm{T}})
    =\Phi_{\mathrm{dist}}(X_{\mathrm{K}}).
\end{equation}
An MLP receiving only $\Phi_{\mathrm{dist}}$ must therefore return the same
output for both members of a pair for every choice of its parameters.  The
distance-MLP probe is nullified at the feature level, rather than merely
expected to perform poorly after training.

\subsection{Finite Benchmark and Generation-Time Certification}

We instantiate 50 pairs.
\begin{equation}
    (h,p)\in\{(6,4),(8,4),\ldots,(104,4)\}.
\end{equation}
The increasing values of $h$ provide 50 distinct discretization sizes; they
are not themselves the reason for histogram equality.  The essential
conditions are $p=4$, even $h$, and the two seam maps defined above.

The generator does not rely solely on this counting argument.  Before writing
each pair, it computes the exact integer-valued features and aborts unless all
of the following equalities hold:
\begin{enumerate}
    \item rank-wise cell counts and Euler characteristic;
    \item vertex-degree profiles;
    \item edge-to-face incidence profiles;
    \item boundary-loop length multisets;
    \item rational and \(\mathbb Z_2\) Betti numbers;
    \item the complete node-to-face distance histogram;
    \item the vertical closed-walk period for every starting column;
    \item stabilized vertex \(1\)-WL colors; and
    \item stabilized rank-colored Hasse-graph \(1\)-WL colors.
\end{enumerate}
For the vertical-period check, both classes have period $2h$ in every column. These
checks are deterministic and are repeated during data generation.  The same
distance histograms and cycle-period signatures are recorded in the
per-pair structural certificates used for auditing the benchmark. A limited
SMCN evaluation of this family is reported below. A separate DIST\_MLP run is
unnecessary because its input vectors are exactly equal within every pair.

\subsection{Original versus Distance-Histogram-Matched Construction}

The topology is unchanged:
\begin{itemize}
    \item \textbf{Class A:} punctured torus;
    \item \textbf{Class B:} punctured Klein bottle;
    \item both have Betti numbers \((1,2,0)\), Euler characteristic \(-1\), one boundary loop of length \(4\), and identical cell counts.
\end{itemize}

\paragraph{Original pairs.}
The original punctured torus used the half-period translation
\[
    (h-1,j)\sim(0,j+p/2 \bmod p),
\]
whereas the punctured Klein bottle used the reflected identification
\[
    (h-1,j)\sim(0,p-1-j \bmod p).
\]
Although these maps give the same vertical period, their seam-displacement
profiles differ. Across the original range of \(h\) and \(p\), this produced
different aggregate node-to-face distance histograms on \(45\) of \(50\)
pairs. Consequently, \texttt{DIST\_MLP} separated \(45/50\) pairs without
establishing orientability detection.

\paragraph{Distance-histogram-matched pairs.}
The revised family retains the half-period translation for the torus and the
reflection for the Klein bottle, but fixes \(p=4\) and restricts \(h\) to even
values. Both seam maps are then fixed-point-free involutions, so their vertical
periods agree:
\[
    \ell_{\mathrm{torus}}
    =
    \ell_{\mathrm{Klein}}
    =
    2h.
\]
The parameter restriction also makes the complete aggregate node-to-face
distance histograms equal, as shown in
Eq.~\eqref{eq:matched-histogram-closed-form}.

The new dataset contains
\[
    p=4,
    \qquad
    h\in\{6,8,10,\ldots,104\},
\]
giving \(50\) pairs. For every pair, the generator verifies the equality of:
\begin{itemize}
    \item numbers of vertices, edges, and faces;
    \item degree profiles;
    \item edge-to-face incidence profiles;
    \item boundary-loop lengths;
    \item rational and \(\mathbb{Z}_2\) Betti numbers;
    \item vertical-cycle periods;
    \item node-to-face distance histograms;
    \item stabilized vertex \(1\)-WL colors; and
    \item stabilized colored Hasse-graph WL colors.
\end{itemize}

\paragraph{Why the models struggle.}
The original dataset allowed models to exploit aggregate metric correlates of
the gluing. In the revised dataset, that distance statistic and the audited
local refinements are matched. GIN cannot distinguish the paired vertex graphs
under the tested \(1\)-WL refinement, and the corresponding rank-colored Hasse
graphs are also \(1\)-WL equivalent under the tested refinement. The intended
orientation distinction is global, but this does not mean that every remaining
signal is topological: all \(50\) pairs have different graph-Laplacian spectra,
so the spectral channel is a measured leak.

The limited SMCN evaluation is consistent with a locality bottleneck but does
not establish one. SMCN separated only the two smallest pairs, \(h=6\) and
\(h=8\), and failed on the next \(29\) completed pairs. After \(31\) pairs had
completed, the run exhausted a \(40\,\mathrm{GB}\) GPU. This pattern is
consistent with small
complexes permitting relevant information to propagate through message passing
or pooling, but it does not identify the causal feature.

One caveat should remain explicit: ``metric-matched'' does \emph{not} prove
that the two finite complexes are isometric or that every conceivable metric
statistic agrees. It means that the specific distance histogram, vertical
period, and audited local combinatorial statistics used by the probes are
matched. This is a stronger control than the original family, but not an
equality theorem for every possible metric invariant.

\subsection{Scope of the Control}

This construction directly addresses the shortcut identified by the
distance probe: \textbf{DIST\_MLP}. The probe receives identical feature vectors on the two
classes.  It also matches the specified boundary and vertical-period
channels.  It does \textbf{not} imply that the two cell complexes are isometric,
that their labeled distance matrices coincide, or that all possible cycle,
spectral, and higher-order incidence statistics agree.  In particular,
per-cell distance profiles, complete cycle-length distributions, Laplacian
spectra, and higher-order neighborhoods remain outside the stated guarantee.
Indeed, the graph-Laplacian spectra differ on all \(50\) pairs, so this family
is distance-histogram matched but not spectrally matched.
This distinction is intentional.  The revised family is a controlled test of
whether a method uses information beyond the documented count, boundary, and
aggregate-distance shortcuts, not a claim that every non-topological signal
has been eliminated.

\providecommand{\code}[1]{\texttt{#1}}
\providecommand{\smcn}{\code{SMCN}}

\section{Depth Sweep for the Punctured Orientability Family}
\label{app:depth-sweep}

This appendix examines how the separation achieved by \textsc{smcn} changes with
message-passing depth on the punctured torus--punctured Klein-bottle family. We
first audit the earliest closed-walk difference in the constructed
$1$-skeletons, then report the depth sweep. The two analyses are related by a
one-sided containment: the audit identifies the earliest measured
closed-walk-count difference, and the sweep is tested against the corresponding
diagnostic scale. Neither
establishes that the audited statistic is the feature the model uses.

% ---------------------------------------------------------------------
\subsection{Construction and Cycle Audit}\label{app:ds-audit}

Both members are obtained from an $h \times p$ rectangular grid. The row
direction is closed by seam edges and faces joining the last row to the first;
we denote the resulting periodic seam adjacency by $\sim$. The members differ
in the seam map:
\begin{align}
  P^{0}_{h,p} &: \quad (h-1,j) \sim \bigl(0,\; j + p/2 \bmod p\bigr), \\
  P^{1}_{h,p} &: \quad (h-1,j) \sim \bigl(0,\; p-1-j \bmod p\bigr),
\end{align}
where $p$ is even. The first map is a translation by $p/2$; the second is a
reflection. Their cyclic seam displacements satisfy
\[
  d_p\bigl(j,\, j + p/2\bigr) = p/2 \quad \text{for all } j,
  \qquad
  \min_{j} d_p\bigl(j,\, p-1-j\bigr) = 1 .
\]
The reflected seam therefore admits shorter seam-crossing closures at some
columns, while the translation displaces every column uniformly.

For each evaluated pair let $A_0, A_1$ be the two $1$-skeleton adjacency
matrices and define
\[
  L^{\star} \;=\; \min\bigl\{\ell \ge 1 :
    \operatorname{tr}(A_0^{\ell}) \neq \operatorname{tr}(A_1^{\ell})\bigr\}.
\]
The audit found $L^{\star} = h+1$ at all $50$ grid points,
$h \in \{4,\dots,13\}$, $p \in \{4,6,8,10,12\}$. Traces were compared in exact
integer arithmetic modulo two large primes, so no floating-point tolerance
enters. This is a reproducible finite-grid check, not a theorem for arbitrary
discretizations or architectures.

\begin{table}[t]
\caption{Closed-walk and typed-refinement audit on the $50$ evaluated grid
points. For every tested $p$ the first differing closed-walk length is
$L^{\star} = h+1$. For $p \ge 6$, rank-colored Hasse $1$-WL first separates at
round $h+2$; for $p = 4$ it does not separate within $400$ rounds. The quantity
$\lceil L^{\star}/2 \rceil$ is a neighborhood-radius diagnostic and is not
asserted to be a sufficient model depth.}
\label{tab:cycle-audit}
\centering\small
\begin{tabular}{@{}rcccc@{}}
\toprule
$h$ & $L^{\star}$ & $\lceil L^{\star}/2 \rceil$
    & Hasse $1$-WL round ($p \ge 6$) & Hasse $1$-WL ($p = 4$) \\
\midrule
 4 &  5 & 3 &  6 & not separated \\
 5 &  6 & 3 &  7 & not separated \\
 6 &  7 & 4 &  8 & not separated \\
 7 &  8 & 4 &  9 & not separated \\
 8 &  9 & 5 & 10 & not separated \\
 9 & 10 & 5 & 11 & not separated \\
10 & 11 & 6 & 12 & not separated \\
11 & 12 & 6 & 13 & not separated \\
12 & 13 & 7 & 14 & not separated \\
13 & 14 & 7 & 15 & not separated \\
\bottomrule
\end{tabular}

\end{table}

As a cross-check we enumerated simple cycles on four small grids. The counts
agree at length $h$ and differ at $h+1$ in every tested case
(Table~\ref{tab:simple-cycles}).

\begin{table}[t]
\caption{Simple-cycle counts on four small grids. This finite cross-check agrees
with the closed-walk audit at lengths $h$ and $h+1$; it does not characterize
simple cycles at larger sizes or lengths.}
\label{tab:simple-cycles}
\centering\small
\begin{tabular}{@{}rrrrr@{}}
\toprule
$h$ & $p$ & cycle length & $P^{0}_{h,p}$ & $P^{1}_{h,p}$ \\
\midrule
4 & 6 & $h$   &  24 &  24 \\
4 & 6 & $h+1$ &   0 &  16 \\
5 & 6 & $h$   &   0 &   0 \\
5 & 6 & $h+1$ &  65 &  85 \\
6 & 6 & $h$   &  78 &  78 \\
6 & 6 & $h+1$ &   0 &  24 \\
7 & 6 & $h$   &   0 &   0 \\
7 & 6 & $h+1$ & 504 & 532 \\
\bottomrule
\end{tabular}

\end{table}

% ---------------------------------------------------------------------
\subsection{Depth Sweep}\label{app:ds-sweep}

We vary the number of subgraph message-passing layers
$L \in \{1,2,3,4,5,6,8\}$ with three model seeds per depth. Every run evaluates
the same $50$ pairs on $h \in \{4,\dots,13\}$, $p \in \{4,6,8,10,12\}$. The
model uses binary face-corner marking, so no explicit shortest-path value is
supplied and all discrimination must be built by message passing; a fresh model
is initialized for every pair, so a pair's outcome cannot depend on its position
in the evaluation stream. Remaining settings follow the saved run
configurations: $32$ permutation copies, $20$ epochs, embedding dimension $32$,
threshold $\tau = 60$.

\paragraph{Provenance.} Depths $L \le 5$ were produced by one evaluator
implementation and $L \in \{6,8\}$ by a second, functionally equivalent one. A
control run of $L = 5$, seed $0$ under the second implementation scored $23/50$
against $24/50$ under the first, a one-pair discrepancy consistent with
floating-point reduction order in the separation statistic. Cross-depth
comparisons in this appendix should be read with that tolerance in mind.

\begin{table}[t]
\caption{Depth-sweep results. Seed columns report pairs satisfying both target
separation and permutation reliability; the all-seeds column counts pairs solved
under every seed. Reliability is $50/50$ for every seed at every depth, so the
zeros elsewhere are failures to separate rather than degenerate models.}
\label{tab:depth-solved}
\centering\small
\begin{tabular}{@{}rcccccc@{}}
\toprule
$L$ & seed $0$ & seed $1$ & seed $2$ & mean & solved by all seeds & reliability \\
\midrule
1 &  0 &  0 &  0 &  $0.0$ &  $0/50$ & $50/50$ \\
2 & 10 & 10 &  9 &  $9.7$ &  $9/50$ & $50/50$ \\
3 & 17 & 16 & 16 & $16.3$ & $13/50$ & $50/50$ \\
4 & 21 & 21 & 22 & $21.3$ & $17/50$ & $50/50$ \\
5 & 24 & 25 & 27 & $25.3$ & $21/50$ & $50/50$ \\
6 & 30 & 28 & 26 & $28.0$ & $24/50$ & $50/50$ \\
8 & 32 & 31 & 34 & $32.3$ & $28/50$ & $50/50$ \\
\bottomrule
\end{tabular}

\end{table}

\begin{table}[t]
\caption{Solved pair--seed evaluations stratified by grid height. Each entry
pools five values of $p$ and three seeds, hence a denominator of $15$. The
``largest $h$'' column is a maximum over those $15$ evaluations and is therefore
supported by as little as one evaluation: the entries at $L=4$ ($h=9$) and
$L=6$ ($h=10$) rest on $1/15$ and $2/15$ respectively. No entry exceeds the
diagnostic scale $2L+1$.}
\label{tab:depth-height}
\centering\small
\setlength{\tabcolsep}{3.5pt}
\begin{tabular}{@{}rcccccccccccc@{}}
\toprule
$L$ & $h{=}4$ & $h{=}5$ & $h{=}6$ & $h{=}7$ & $h{=}8$ & $h{=}9$ & $h{=}10$
    & $h{=}11$ & $h{=}12$ & $h{=}13$ & largest $h$ & scale $2L{+}1$ \\
\midrule
1 &  $0/15$ &  $0/15$ &  $0/15$ &  $0/15$ &  $0/15$ &  $0/15$ & $0/15$ & $0/15$ & $0/15$ & $0/15$ & none & $3$ \\
2 & $15/15$ & $14/15$ &  $0/15$ &  $0/15$ &  $0/15$ &  $0/15$ & $0/15$ & $0/15$ & $0/15$ & $0/15$ & $5$  & $5$ \\
3 & $15/15$ & $15/15$ & $13/15$ &  $6/15$ &  $0/15$ &  $0/15$ & $0/15$ & $0/15$ & $0/15$ & $0/15$ & $7$  & $7$ \\
4 & $15/15$ & $15/15$ & $13/15$ & $14/15$ &  $6/15$ &  $1/15$ & $0/15$ & $0/15$ & $0/15$ & $0/15$ & $9$  & $9$ \\
5 & $15/15$ & $15/15$ & $15/15$ & $15/15$ &  $9/15$ &  $7/15$ & $0/15$ & $0/15$ & $0/15$ & $0/15$ & $9$  & $11$ \\
6 & $15/15$ & $15/15$ & $15/15$ & $15/15$ & $13/15$ &  $9/15$ & $2/15$ & $0/15$ & $0/15$ & $0/15$ & $10$ & $13$ \\
8 & $15/15$ & $15/15$ & $15/15$ & $15/15$ & $15/15$ & $12/15$ & $4/15$ & $4/15$ & $2/15$ & $0/15$ & $12$ & $17$ \\
\bottomrule
\end{tabular}

\end{table}

\begin{table}[t]
\caption{Median target $T^2$ over the $15$ pair--seed evaluations at each height;
the threshold is $\tau = 60$. The final column is the largest of the four
height-wise medians for $h = 10,\dots,13$. These are diagnostic margins, not
accuracies, and should be read with Table~\ref{tab:depth-solved}.}
\label{tab:depth-t2}
\centering\scriptsize
\setlength{\tabcolsep}{2.5pt}
\begin{tabular}{@{}rrrrrrrr@{}}
\toprule
$L$ & $h{=}4$ & $h{=}5$ & $h{=}6$ & $h{=}7$ & $h{=}8$ & $h{=}9$
    & max.\ median, $h{=}10{:}13$ \\
\midrule
1 & $5.68\times10^{-7}$ & $1.74\times10^{-6}$ & $8.88\times10^{-7}$ & $2.27\times10^{-6}$ & $1.72\times10^{-5}$ & $1.42\times10^{-5}$ & $1.11\times10^{-4}$ \\
2 & $4.51\times10^{5}$  & $7.29\times10^{3}$  & $1.32\times10^{-4}$ & $2.25\times10^{-4}$ & $1.42\times10^{-4}$ & $7.16\times10^{-4}$ & $2.15\times10^{-3}$ \\
3 & $3.63\times10^{6}$  & $1.73\times10^{5}$  & $2.27\times10^{3}$  & $2.48\times10^{1}$  & $1.80\times10^{-3}$ & $7.72\times10^{-3}$ & $1.20\times10^{-2}$ \\
4 & $5.69\times10^{7}$  & $2.86\times10^{6}$  & $2.56\times10^{4}$  & $5.04\times10^{3}$  & $5.96$              & $3.05\times10^{-2}$ & $3.71\times10^{-2}$ \\
5 & $6.75\times10^{7}$  & $2.97\times10^{7}$  & $7.27\times10^{5}$  & $1.05\times10^{4}$  & $6.90\times10^{2}$  & $2.64\times10^{1}$  & $1.90\times10^{-1}$ \\
6 & $1.81\times10^{8}$  & $2.30\times10^{7}$  & $1.07\times10^{6}$  & $1.00\times10^{5}$  & $5.62\times10^{3}$  & $1.51\times10^{2}$  & $9.69\times10^{-1}$ \\
8 & $9.23\times10^{8}$  & $3.93\times10^{7}$  & $3.89\times10^{6}$  & $1.96\times10^{5}$  & $3.09\times10^{3}$  & $1.99\times10^{2}$  & $4.04$ \\
\bottomrule
\end{tabular}

\end{table}

\paragraph{Containment below the audited scale.}
The robust statement available from this sweep is one-sided. Across the $450$
pair--seed evaluations in Table~\ref{tab:depth-height} with $h > 2L+1$, not one
was separated. This survives the most permissive possible criterion: a single
separating evaluation out of fifteen would count as a violation, and none
occurred at any depth or seed. An independent sweep on a second punctured family
with a different grid adds a further $267$ evaluations above the scale, also
with no separations.

The converse does not hold, and we do not claim it. The largest separated height
does not track $2L+1$: the two coincide at $L = 2,3,4$, but the $L=4$ entry rests
on a single evaluation of fifteen, and from $L = 5$ onward the observed height
falls increasingly short of the scale ($9$ against $11$, $10$ against $13$, $12$
against $17$). The scale is therefore informative at small depth and
progressively loose as $L$ grows.

\paragraph{The trend does not saturate.}
Both aggregate measures increase monotonically across the swept range: the mean
solved count rises $0.0 \to 9.7 \to 16.3 \to 21.3 \to 25.3 \to 28.0 \to 32.3$,
and the largest separated height rises from none to $12$. The apparent plateau
between $L = 4$ and $L = 5$, where both give $h = 9$, does not persist once
$L = 6$ and $L = 8$ are included. Table~\ref{tab:depth-t2} shows the associated
margin pattern: target medians generally increase with depth at the smaller
heights, with some nonmonotonic variation across depths, and decrease sharply
as $h$ grows. Consequently, the threshold crossing moves to larger heights
only gradually.

% ---------------------------------------------------------------------
\subsection{Interpretation and Limitations}\label{app:ds-interpretation}

Across $L = 1$ to $L = 8$ the mean solved count rises from $0.0/50$ to
$32.3/50$, and separation remains confined below the audited cycle scale in all
$450$ evaluations where the comparison applies. These observations establish
an empirical depth-dependent separation trend for this model, representation,
training budget, and finite grid. The observed separations remain below the
audited scale, but this containment is not a general depth bound.

They do not establish a depth barrier, and they do not identify cycles as the
causal feature. Depth varies receptive field and model capacity together, and
this sweep contains no width-matched control, so a capacity explanation of the
same curve is not excluded. Equality of selected closed-walk counts does not
imply equality of all radius-$L$ neighborhoods, and the complexes may retain
other realization-level signals. Reliability being $50/50$ throughout means the
permutation control never fires, so it certifies stability but discriminates
nothing.

The saved evaluator warns that pairs with $h < 5$ or $p < 5$ place the puncture
and the twist seam in adjacent neighborhoods, so those pairs may admit a more
local cue, and the sweep includes $p = 4$. Two observations bound this concern.
For $L\in\{2,3,4,5\}$, solved rates at $p = 4$ track those at $p \ge 6$ closely
($17\%$ against $20\%$ at $L=2$, $33\%$ against $32\%$ at $L=3$, $43\%$ against
$42\%$ at $L=4$, $47\%$ against $52\%$ at $L=5$), and restricting the sweep to
$p \ge 6$ leaves the largest separated height unchanged at every depth. Further,
$p = 4$ is the one column in which rank-colored Hasse $1$-WL does not separate
the pair within the audited $400$ rounds (Table~\ref{tab:cycle-audit}). Thus,
the tested typed-refinement audit yields no shortcut exactly where the
local-cue caveat applies; this is not a proof of equivalence under every
refinement variant.

We therefore describe the result as an empirical depth-dependent separation
trend aligned with an audited cycle scale, and not as proof of a global cycle
barrier or of orientability recognition.

Three extensions would sharpen this. A width-matched shallow control would
separate depth from capacity and is the single most valuable addition. A
marking-radius ablation, $m_k(v,c) = \min(d(v,c), k)$ for
$k \in \{1,2,3,\infty\}$, interpolates between the binary marking used here and
the shortest-path marking of the main results, and would separate reachability of
the signal from recoverability of it. Establishing whether the observed
containment persists under matched-capacity models and broader grid families
would require a separate theoretical or substantially expanded empirical
analysis; we leave this as an open problem rather than a conjecture.

% ============================================================
\section{Fundamental Group Benchmark Construction}
\label{app:fg-construction}
% ============================================================

The Fundamental Group Family is designed to compare two complexes with
identical integral homology but nonisomorphic fundamental groups. Both members
are constructed from four generator loops and one attached \(2\)-cell.

\subsection{General Presentation Complex}
\label{app:fg-presentation}

Let
\[
F_4=\langle a,b,c,d\rangle
\]
be the free group on four generators. For a word \(w\in F_4\), define
\[
X_w
=
\left(\bigvee\nolimits^4 S^1\right)\cup_w e^2.
\]
The space \(\bigvee^4S^1\) has one vertex and one loop for each generator.
The boundary of the \(2\)-cell is attached along the loop described by \(w\).

By the Seifert--van Kampen theorem,
\[
\pi_1(X_w)
\cong
\langle a,b,c,d\mid w\rangle
\]
~\citep{hatcher2002algebraic,lyndon1977combinatorial}.

In the emitted combinatorial complex, each generator loop is subdivided into
an edge path and the attaching polygon is subdivided consistently. The
certifier retains the ordered signed word \(w\). The machine representation
\(\Phi(X_w)\), however, contains only the resulting unsigned incidence,
adjacency, and bridge relations.

\subsection{The Surface Group Member}
\label{app:fg-surface-member}

For group elements \(x\) and \(y\), their commutator is
\[
[x,y]=xyx^{-1}y^{-1}.
\]
The canonical genus \(2\) surface relator is
\[
w_S=[a,b][c,d]
=
aba^{-1}b^{-1}cdc^{-1}d^{-1}.
\]
The corresponding presentation is
\[
G_S
=
\langle a,b,c,d\mid[a,b][c,d]\rangle.
\]
This is the fundamental group of the closed orientable surface
\(\Sigma_2\) of genus \(2\).

The benchmark generator may insert freely cancelling subwords into \(w_S\).
Every accepted surface word nevertheless satisfies
\[
\operatorname{red}(w_S^{\mathrm{noise}})
=
[a,b][c,d],
\]
where \(\operatorname{red}\) denotes free reduction. Since freely equal words
describe homotopic loops in \(\bigvee^4S^1\), attaching along the noisy word
produces a presentation complex homotopy equivalent to the canonical surface
presentation complex. Therefore,
\[
\pi_1(X_{w_S})\cong G_S.
\]

\subsection{The Free Group Member}
\label{app:fg-free-member}

The second member uses a word \(w_F\) that freely reduces to the identity:
\[
\operatorname{red}(w_F)=1.
\]
Its presentation is
\[
G_F
=
\langle a,b,c,d\mid w_F\rangle
\cong
\langle a,b,c,d\rangle
=
F_4.
\]
Because \(w_F\) represents a null-homotopic loop in \(\bigvee^4S^1\), attaching
the \(2\)-cell along \(w_F\) imposes no relation on the fundamental group.
Topologically,
\[
X_{w_F}
\simeq
\left(\bigvee\nolimits^4S^1\right)\vee S^2.
\]
The extra \(S^2\) appears because a \(2\)-cell attached along a null-homotopic
map contributes a spherical summand.

\subsection{Why the Fundamental Groups Differ}
\label{app:fg-group-difference}

The two groups are
\[
G_S
=
\langle a,b,c,d\mid[a,b][c,d]\rangle,
\qquad
G_F=F_4.
\]
They are not isomorphic. One way to see this is through group homology. The
surface \(\Sigma_2\) is an Eilenberg--MacLane space \(K(G_S,1)\), so
\[
H_2(G_S;\mathbb Z)
\cong
H_2(\Sigma_2;\mathbb Z)
\cong
\mathbb Z.
\]
A wedge of four circles is a \(K(F_4,1)\), and therefore
\[
H_2(F_4;\mathbb Z)=0.
\]
Since isomorphic groups have isomorphic group homology,
\[
G_S\not\cong F_4.
\]
Equivalently, \(G_S\) is a closed surface group of cohomological dimension
\(2\), whereas \(F_4\) is a free group of cohomological dimension \(1\).

\subsection{Why the Cellular Homology Is Identical}
\label{app:fg-homology}

The presentation complex has one \(0\)-cell, four \(1\)-cells, and one
\(2\)-cell. Its cellular chain complex is
\[
0
\longrightarrow
\mathbb Z
\xrightarrow{\partial_2}
\mathbb Z^4
\xrightarrow{\partial_1}
\mathbb Z
\longrightarrow
0.
\]
Since every generator loop begins and ends at the same vertex,
\[
\partial_1=0.
\]

For a word \(w\), let \(\epsilon_x(w)\) denote the total signed exponent of
the generator \(x\). The two-dimensional boundary map is
\[
\partial_2(1)
=
\begin{pmatrix}
\epsilon_a(w)\\
\epsilon_b(w)\\
\epsilon_c(w)\\
\epsilon_d(w)
\end{pmatrix}.
\]
The surface relator is a product of commutators, so every exponent sum is zero:
\[
\epsilon_a(w_S)
=
\epsilon_b(w_S)
=
\epsilon_c(w_S)
=
\epsilon_d(w_S)
=
0.
\]
The free word reduces to the identity and also has zero exponent sums.
Therefore,
\[
\partial_2=0
\]
for both presentation complexes. It follows that
\[
H_0(X_{w_S};\mathbb Z)
\cong
H_0(X_{w_F};\mathbb Z)
\cong
\mathbb Z,
\]
\[
H_1(X_{w_S};\mathbb Z)
\cong
H_1(X_{w_F};\mathbb Z)
\cong
\mathbb Z^4,
\]
and
\[
H_2(X_{w_S};\mathbb Z)
\cong
H_2(X_{w_F};\mathbb Z)
\cong
\mathbb Z.
\]
Thus,
\[
H_*(X_{w_S};\mathbb Z)
\cong
H_*(X_{w_F};\mathbb Z)
\cong
(\mathbb Z,\mathbb Z^4,\mathbb Z).
\]

The \(H_2\) classes have different origins. In the surface member, the class
is the fundamental class of the genus \(2\) surface. In the free member, it
is represented by the \(S^2\) summand. Cellular homology records that both
classes exist, but it does not record this difference in their homotopical
origin.

\subsection{Controlled Word Generation}
\label{app:fg-word-generation}

Noise is added by inserting freely cancelling subwords. If
\[
w=uv
\]
and \(x\in\{a^{\pm1},b^{\pm1},c^{\pm1},d^{\pm1}\}\), then
\[
w'
=
uxx^{-1}v
\]
has the same free reduction as \(w\). More general neutral words of the form
\(zz^{-1}\) may be inserted in the same way:
\[
\operatorname{red}(uzz^{-1}v)
=
\operatorname{red}(uv).
\]

These insertions change the length and local combinatorics of the attaching
polygon without changing the represented group. To prevent the two classes
from being separated by simple properties of their words, the rejection
sampler matches:

\begin{enumerate}
    \item total attaching-word length;
    \item signed and unsigned generator frequencies;
    \item the number of immediate cancellations;
    \item the number of local backtracking motifs; and
    \item the resulting cell counts after subdivision.
\end{enumerate}

A candidate pair is rejected if any required statistic differs.

\subsection{Generator Gates}
\label{app:fg-gates}

Every accepted pair passes the following checks:

\begin{enumerate}
    \item \textbf{Target validity.}
    The surface word freely reduces to \([a,b][c,d]\), while the free word
    reduces to the identity.

    \item \textbf{Cell counts.}
    The numbers of rank \(0\), rank \(1\), and rank \(2\) cells agree.

    \item \textbf{Integral homology.}
    Smith normal form verifies
    \[
    H_*=(\mathbb Z,\mathbb Z^4,\mathbb Z)
    \]
    without torsion for both members
    ~\citep{kannan1979polynomial,storjohann1996near}.

    \item \textbf{Local incidence.}
    Vertex degrees, edge-to-face incidence degrees, and face-size
    distributions agree.

    \item \textbf{Vertex links.}
    The multisets of vertex-link profiles agree.

    \item \textbf{Boundary structure.}
    Both complexes have zero boundary edges.

    \item \textbf{Hasse refinement.}
    Rank-colored Hasse \(1\)-WL produces matching color partitions through
    stabilization. Stabilization occurs at round \(6\) for the evaluated
    family.

    \item \textbf{Word statistics.}
    Word length, generator frequencies, immediate cancellations, and
    backtracks agree.
\end{enumerate}

These gates control the count, homology, local incidence, boundary, word, and
tested refinement channels. The restricted count, boundary, and distance
probes each obtain \(0/64\) in the final evaluation. The distance result shows
that the tested shortest-path histogram is not sufficient by itself. It does
not prove that every possible metric statistic is absent.

\subsection{Representation-Level Limitation}
\label{app:fg-representation-limit}

The mathematical construction begins with an ordered signed word. For example,
the letters \(a\) and \(a^{-1}\) describe traversal of the same generator in
opposite directions, and the order of the letters determines the attaching
loop. The emitted representation \(\Phi(X_w)\) does not directly provide this
ordered signed sequence to the tested models. It contains the unsigned
incidence and adjacency relations produced after realization.

Consequently, the benchmark establishes that the two mathematical complexes
have different fundamental groups, but a positive model result does not prove
that the model reconstructed the original attaching word. The model may
instead detect another global structural consequence of the realization. This
is why the resulting claim remains representation-relative and certificate-qualified.

\subsection{Dataset Size and Scope}
\label{app:fg-scope}

The rejection sampler accepts approximately one candidate in \(57\). The final
dataset contains \(16\) certified genus-\(2\) pairs. Evaluating four model seeds
produces
\[
16\times4=64
\]
pair--seed evaluations.

The accepted complexes occupy three principal word-length and cell-count
groups, containing approximately \(48\), \(72\), and \(120\) cells. The
generator can be extended to genus \(g\) by using
\[
w_g
=
\prod_{i=1}^{g}[a_i,b_i]
\]
on \(2g\) generators. The corresponding comparison would be
\[
\left\langle
a_1,b_1,\ldots,a_g,b_g
\ \middle|\
\prod_{i=1}^{g}[a_i,b_i]
\right\rangle
\qquad\text{versus}\qquad
F_{2g},
\]
with matched cellular homology
\[
H_*=(\mathbb Z,\mathbb Z^{2g},\mathbb Z).
\]
Only the genus \(2\) family is certified and evaluated in the present study.
No claim is made for higher genus, unseen presentations, or arbitrary groups.

\begin{table*}[t]
\caption{Topological properties of the four benchmark pairs. Entries follow the order of the spaces in the first column. The symbol \(g\) denotes orientable genus, \(k\) denotes crosscap number, and \(b\) denotes the number of boundary components.}
\label{tab:topology}
\centering
\scriptsize
\setlength{\tabcolsep}{4pt}
\renewcommand{\arraystretch}{1.12}

\begin{tabular}{@{}llll@{}}
\toprule
\textbf{Pair} &
\textbf{Integral homology} &
\textbf{Homotopy} &
\textbf{Genus, crosscaps, and boundary}\\
\midrule

\shortstack[l]{\textbf{Cyl. versus M\"ob.}\\\textbf{Orientability}} &
Both \((\mathbb Z,\mathbb Z,0)\) &
Both \(\simeq S^1\) &
\(g=0,b=2\) versus \(k=1,b=1\)\\

\shortstack[l]{\textbf{Tor. versus Klein}\\\textbf{Orientability}} &
\shortstack[l]{\((\mathbb Z,\mathbb Z^2,\mathbb Z)\) versus\\
\((\mathbb Z,\mathbb Z\oplus\mathbb Z_2,0)\)} &
Not homotopy equivalent &
\(g=1,b=0\) versus \(k=2,b=0\)\\

\shortstack[l]{\textbf{Punctured torus versus punctured Klein bottle}\\\textbf{Orientability}} &
Both \((\mathbb Z,\mathbb Z^2,0)\) &
Both \(\simeq S^1\vee S^1\) &
\(g=1,b=1\) versus \(k=2,b=1\)\\

\shortstack[l]{\textbf{Surf. versus free group}\\\textbf{Fundamental Group}} &
Both \((\mathbb Z,\mathbb Z^4,\mathbb Z)\) &
\shortstack[l]{\(\Sigma_2\) versus\\
\((\bigvee^4 S^1)\vee S^2\)} &
\(g=2,b=0\) versus non-surface\\

\bottomrule
\end{tabular}

\end{table*}

\begin{table*}[t]
\caption{Mathematical distinction and controlled information in the
genus-\(2\) Fundamental Group Family.}
\label{tab:app-fg-members}
\centering
\small
\begin{tabular}{@{}lll@{}}
\toprule
\textbf{Property}
& \textbf{Surface member}
& \textbf{Free member}\\
\midrule
Reduced attaching word
& \([a,b][c,d]\)
& Identity\\

Fundamental group
& \(\langle a,b,c,d\mid[a,b][c,d]\rangle\)
& \(F_4\)\\

Homotopy type
& Genus \(2\) surface
& \((\bigvee^4S^1)\vee S^2\)\\

Integral homology
& \((\mathbb Z,\mathbb Z^4,\mathbb Z)\)
& \((\mathbb Z,\mathbb Z^4,\mathbb Z)\)\\

Cell counts
& Matched
& Matched\\

Local incidence and vertex links
& Matched
& Matched\\

Boundary
& None
& None\\

Rank-colored Hasse \(1\)-WL
& Matched through stabilization
& Matched through stabilization\\

Word statistics
& Matched
& Matched\\

Restricted distance probe
& \multicolumn{2}{c}{No separation in \(64\) evaluations}\\
\bottomrule
\end{tabular}

\end{table*}

\section{A Held-Out Relator Family}\label{app:relator}

The Fundamental Group experiment in the main text rests on a single
presentation family: the genus-$2$ surface relator against a freely trivial
relator. To test that the reported separation is not specific to that one
construction, we add a second family built from a \emph{different relator},
which played no role in designing or tuning the generator.

% ---------------------------------------------------------------------------
\subsection{Construction}\label{app:relator:construction}

Both members of a pair are built identically up to the attaching word. Take a
wedge of $2g=4$ circles, subdivide each into a path of length $p=3$, and attach
a single triangulated $2$-cell along a closed edge-path $w$. The pair differs
only in which word is used:
\[
  \text{member A: } w_A \xrightarrow{\ \text{free reduction}\ }
  \bigl[[a_1,b_1],[a_2,b_2]\bigr],
  \qquad
  \text{member B: } w_B \xrightarrow{\ \text{free reduction}\ } 1 .
\]
By van Kampen, $\pi_1$ of the resulting complex is the free group $F_4$ modulo
the normal closure of the reduced relator. Member B's $2$-cell is attached along
a null-homotopic loop and therefore imposes no relation.

\paragraph{Why homology cannot separate them.}
Both relators lie in the commutator subgroup $[F,F]$, so both abelianize to $0$;
the cellular boundary $\partial_2$ vanishes and
\begin{equation}\label{eq:ff}
  r\in[F,F] \;\Longrightarrow\;
  H_\ast \;=\; \bigl(\mathbb{Z},\,\mathbb{Z}^{2g},\,\mathbb{Z}\bigr),
  \ \text{torsion-free.}
\end{equation}
Homology matching is thus automatic rather than searched for; we nonetheless
recompute it exactly (Smith normal form) on every emitted pair. A second
consequence of \eqref{eq:ff} is used in \S\ref{app:relator:budget}: the exponent
sum of each generator vanishes, so every generator occurs an \emph{even} number
of times.

\begin{table}[H]\centering\small
\caption{The held-out pair. Both members are presentation complexes over the
same wedge of four circles at the same subdivision, differing only in the
attaching word. All listed generator-gated quantities---cell counts,
homology, local incidence, and $1$-WL color refinement---are matched by
construction or rejection; the fundamental group is the only intended
difference.}\label{tab:pair}
\begin{tabularx}{\linewidth}{@{}>{\raggedright\arraybackslash}p{0.26\linewidth}>{\raggedright\arraybackslash}X>{\raggedright\arraybackslash}X@{}}
\toprule
 & \textbf{Member A} & \textbf{Member B}\\
\midrule
Relator (freely reduced) & $\bigl[[a_1,b_1],[a_2,b_2]\bigr]$ & $1$\\
$\pi_1$ & $\langle a_1,b_1,a_2,b_2 \mid [[a_1,b_1],[a_2,b_2]]\rangle$ & $F_4$\\
Relator length & $16$ & $0$\\
Generator occurrences & $(4,4,4,4)$ & $(0,0,0,0)$\\
Realization of the word & base relator $+$ spliced $xx^{-1}$ pairs & random non-crossing (Dyck) pairing\\
\midrule
\multicolumn{3}{@{}p{\linewidth}@{}}{\emph{After budgeting (\S\ref{app:relator:budget}), the two members are equalized on:}}\\
Word length, occurrences, exponent sums & \multicolumn{2}{>{\raggedright\arraybackslash}p{0.68\linewidth}@{}}{identical}\\
Cell counts, $\chi$, integral homology & \multicolumn{2}{>{\raggedright\arraybackslash}p{0.68\linewidth}@{}}{identical; Betti vector $(1,4,1)$, torsion-free; $\chi=-2$}\\
Local incidence, vertex-link WL & \multicolumn{2}{>{\raggedright\arraybackslash}p{0.68\linewidth}@{}}{identical}\\
Rank-colored Hasse $1$-WL & \multicolumn{2}{>{\raggedright\arraybackslash}p{0.68\linewidth}@{}}{indistinguishable to stabilization}\\
\midrule
\textbf{Sole intended difference} & \multicolumn{2}{>{\raggedright\arraybackslash}p{0.68\linewidth}@{}}{$\pi_1$, certified non-isomorphic (Table~\ref{tab:hom})}\\
\bottomrule
\end{tabularx}

\end{table}

% ---------------------------------------------------------------------------
\subsection{Certifying that the Groups Differ}\label{app:relator:hom}

Because homology is identical, distinctness of the two groups requires a
separate argument. For a finite group $H$, $N_H(G)=|\operatorname{Hom}(G,H)|$ is
an isomorphism invariant, and for a one-relator group it is computable by
exhaustive evaluation over $H^{4}$. Hence
$N_H(G_1)\neq N_H(G_2)$ for any single $H$ \emph{proves} $G_1\not\cong G_2$. The
converse does not hold, and we report equal counts as ``not separated by this
panel'' rather than as evidence of isomorphism.

\begin{table}[H]\centering\small
\caption{Exact homomorphism counts. The held-out pair is certified
non-isomorphic by $S_4$ alone ($172{,}800$ vs.\ $331{,}776$); the two groups
agree on the four smaller test groups, making this a deliberately close pair
rather than an easy one. The counting routine is validated against two closed
forms: $|\operatorname{Hom}(F_4,H)|=|H|^4$ reproduces the first row exactly, and
the Frobenius--Mednykh formula $|H|^{3}\sum_\chi \chi(1)^{-2}$ reproduces the
surface-group row exactly ($486$, $2{,}176$, $2{,}176$, $5{,}376$,
$34{,}176$).}\label{tab:hom}
\begin{tabular}{@{}lrrrrr@{}}
\toprule
\textbf{Group} & $S_3$ & $D_4$ & $Q_8$ & $A_4$ & $S_4$\\
\midrule
$F_4$ \ (member B)      & $1{,}296$ & $4{,}096$ & $4{,}096$ & $20{,}736$ & $331{,}776$\\
$\langle a,b,c,d \mid [[a,b],[c,d]]\rangle$ \ (member A)
                        & $1{,}296$ & $4{,}096$ & $4{,}096$ & $20{,}736$ & $\mathbf{172{,}800}$\\
\midrule
Genus-$2$ surface group \emph{(original family, for reference)}
                        & $486$     & $2{,}176$ & $2{,}176$ & $5{,}376$  & $34{,}176$\\
\bottomrule
\end{tabular}

\end{table}

% ---------------------------------------------------------------------------
\subsection{Length and Occurrence Budgeting}\label{app:relator:budget}

The two relators have different lengths ($16$ and $0$) and different occurrence
profiles, so pairing them directly would violate the word-length and
per-generator count gates. Both members are therefore filled to a common target
\[
  t_k=\max\bigl(c_A[k],c_B[k]\bigr),\qquad \tau_k=t_k+2N_k ,
\]
where $N$ is a \emph{shared} noise multiset, identical for both members, whose
per-generator multiplicities are $N_k$. Every deficit $\tau_k-c[k]$ is even by
\eqref{eq:ff}, so the budget is always realizable with $xx^{-1}$ pairs; both
members end with identical length $\sum_k\tau_k$ unconditionally, not by search.

The two realization modes in Table~\ref{tab:pair} differ deliberately. Member A
splices $xx^{-1}$ pairs into the base relator at independently drawn positions;
free reduction returns the base relator for any positions, so $\pi_1$ is fixed by
construction. Member B uses a random non-crossing (Dyck) pairing rather than
adjacent $xx^{-1}$ insertions, which would be trivially detectable by backtrack
counting. Residual backtrack differences are removed by rejection-sampling
member B against member A, widening the shared budget and rebuilding both
members on failure, so length and count matching are never put at risk.

All gate functions are imported from the genus-$2$ generator rather than
reimplemented, so the held-out family passes the same stack by construction of
the code, not by assertion.

\begin{table}[H]\centering\small
\caption{Gate stack on an eight-pair audited subset of the 16-pair held-out evaluation family,
with every invariant recomputed from the emitted complexes rather than read from
the stored certificate. Realized complexes occupy cell counts
$(82,300,216)$ and $(106,396,288)$ with attaching-word lengths $72$ and $96$ ---
regimes disjoint from the original family --- and the joint Hasse $1$-WL
partition stabilizes at round $6$ on every pair. Acceptance costs $14.9$
attempts per certified pair (min.\ $3$, max.\ $50$), against roughly $57$ for the
original family.}\label{tab:gates}
\begin{tabular}{@{}llc@{}}
\toprule
\textbf{Level} & \textbf{Gate (exact equality)} & \textbf{Passed}\\
\midrule
Algebraic & Member A freely reduces to $[[a_1,b_1],[a_2,b_2]]$ & $8/8$\\
Algebraic & Member B freely reduces to the identity            & $8/8$\\
Word      & Length; slot multiset; occurrences; exponent sums   & $8/8$\\
Word      & Realized node-token multiset; backtrack count       & $8/8$\\
Complex   & Cell counts; Euler characteristic $\chi=-2$         & $8/8$\\
Complex   & Betti vector $(1,4,1)$, torsion-free (SNF)     & $8/8$\\
Local     & Local incidence profile; vertex-link WL profile     & $8/8$\\
Global    & Rank-colored Hasse $1$-WL to stabilization         & $8/8$\\
\bottomrule
\end{tabular}

\end{table}

% ---------------------------------------------------------------------------
\subsection{Results}\label{app:relator:eval}

\begin{table}[H]\centering\small
\caption{BREC-protocol results on the held-out family: $16$ certified pairs,
$32$ permuted copies per class, reliability passing at $100\%$ for every model.
\textsc{smcn} with binary marking scores $11,12,12,12$ across four seeds
($47/64$, $73.4\%$), closely matching its $48/64$ on the original family, while
every model bounded by $1$-WL separates none of the evaluated pairs on a family that played no part
in designing the generator. Seed counts differ by row, so these rows should
not be interpreted as equal-budget comparisons. The \textsc{spec\_mlp} row is included for the opposite
reason: it shows the spectral channel remains exploitable here
(\S\ref{app:relator:limits}).}\label{tab:eval}
\begin{tabular}{@{}llccc@{}}
\toprule
\textbf{Model} & \textbf{Information consumed} & \textbf{Seeds} &
\textbf{Held-out family} & \textbf{Original family}\\
\midrule
\textsc{count\_mlp}    & cell counts                     & $1$ & $0/16$ & $0/64$\\
\textsc{dist\_mlp}     & node-to-cell distance histogram & $1$ & $0/16$ & $0/64$\\
\textsc{boundary\_mlp} & boundary-loop summary           & $1$ & $0/16$ & $0/64$\\
\textsc{gin}           & $1$-skeleton, $1$-WL bounded    & $1$ & $0/16$ & $0/64$\\
\textsc{hasse\_gin}    & Hasse graph, $1$-WL bounded     & $1$ & $0/16$ & $0/64$\\
\textsc{homp/cin}      & higher-order message passing    & $1$ & $0/16$ & $0/64$\\
\midrule
\textsc{smcn} (binary marking) & incidence only, no distances & $4$ &
  $\mathbf{47/64}$ & $48/64$\\
\midrule
\textsc{spec\_mlp}     & eigenvalue summaries only       & $2$ & $25/32$ & $28/32$\\
\bottomrule
\end{tabular}

\end{table}

% ---------------------------------------------------------------------------
\subsection{Scope}\label{app:relator:limits}

This extension addresses generality. The separation reported in the main text is
not confined to one hand-built family: it reproduces at nearly the same rate on a
family with a different relator, a different word-length regime and a group
certified non-isomorphic to both original classes, while every $1$-WL-bounded
architecture separates none of the evaluated pairs.

It does not address channel leakage, and we do not present it as if it did.
Measured on the held-out family, the global spectral channel is matched on $0/8$
pairs and the fine-grained metric sub-channel $C_{5.7}$ on $3/8$ --- the same
leak profile as the original family --- and \textsc{spec\_mlp} accordingly
attains $25/32$. Broadening the set of presentation families strengthens the
evidence for generality while leaving open which channel a successful model
exploits. All presentation families reported here use four generators, as in
the genus-$2$ benchmark setting; higher-genus constructions require a separate
leakage audit and are not claimed.

\section{Experimental and Sanity Check Details}
\label{app:experimental-details}

All cells of the same rank begin with constant features. GIN receives constant
vertex features, Hasse GIN receives cell rank labels, and the cellular models
receive constant rank-wise embeddings. SMCN additionally receives its
subcomplex marking. Binary marking indicates whether a vertex is incident to
the marked face, whereas shortest-path marking gives the distance from the
vertex to that face.

Each base complex is expanded into independently permuted copies, with every
incidence, adjacency, and bridge tensor relabeled consistently. The
reliability arm compares two such copies of the same complex. A model that
separates them is permutation unstable, and the corresponding target pair is
not counted as solved.

\subsection{Torus versus Sphere Sanity Experiment}
\label{app:torus-sphere-sanity}

The sanity experiment compares tori and spheres because they differ in several simple and well understood properties. The torus is constructed from a periodic rectangular grid with
\[
(h,p)\in\{(4,4),(5,5),(6,6),(7,7)\}.
\]
For a grid of size \((h,p)\), its cell counts are
\[
(|X_0|,|X_1|,|X_2|)=(hp,2hp,hp).
\]
The sphere is constructed from an icosahedron and its subdivisions. The original icosahedron has \(12\) vertices, \(30\) edges, and \(20\) triangular faces. Subdivision changes the number of cells while preserving the topology of \(S^2\).

The two classes differ in Euler characteristic, homology, and fundamental group:
\[
\chi(T^2)=0,
\qquad
H_*(T^2;\mathbb Z)=(\mathbb Z,\mathbb Z^2,\mathbb Z),
\qquad
\pi_1(T^2)=\mathbb Z^2,
\]
whereas
\[
\chi(S^2)=2,
\qquad
H_*(S^2;\mathbb Z)=(\mathbb Z,0,\mathbb Z),
\qquad
\pi_1(S^2)=\{e\}.
\]
Both spaces are closed, connected, and orientable, so neither has a boundary. Consequently, count and structural models should easily distinguish them, while \textbf{BOUNDARY\_MLP} should show no separation because its input is identical for both classes.

The dataset contains four torus grids and four sphere constructions. Each base complex is expanded into \(32\) independently permuted copies, producing
\[
8\times32=256
\]
samples, with \(128\) samples per class. The data are divided using a stratified \(60/20/20\) training, validation, and test split. This deliberately easy experiment checks whether the complete pipeline detects an available signal and correctly returns a negative result when a model receives no distinguishing information.

\begin{table}[H]
\caption{Supervised torus versus sphere sanity experiment with \(256\) samples
and seed \(0\).}
\label{tab:app-sanity}
\centering
\scriptsize
\setlength{\tabcolsep}{3pt}
\begin{tabular}{@{}lrrrr@{}}
\toprule
\textbf{Model} & \textbf{Parameters} & \textbf{Train accuracy} &
\textbf{Validation accuracy} & \textbf{Test accuracy}\\
\midrule
\texttt{COUNT\_MLP}    & \(226\)      & \(1.000\) & \(1.000\) & \(1.000\)\\
\texttt{DIST\_MLP}     & \(482\)      & \(1.000\) & \(1.000\) & \(1.000\)\\
\texttt{BOUNDARY\_MLP} & \(226\)      & \(0.500\) & \(0.500\) & \(0.500\)\\
\texttt{GIN}           & \(34{,}574\) & \(1.000\) & \(1.000\) & \(1.000\)\\
\texttt{HASSE\_GIN}    & \(34{,}638\) & \(1.000\) & \(1.000\) & \(1.000\)\\
\texttt{HOMP/CIN}      & \(221{,}314\)& \(1.000\) & \(1.000\) & \(1.000\)\\
\bottomrule
\end{tabular}

\end{table}

The sanity experiment verifies the data, training, and evaluation pipeline
before the controlled pair experiments. It uses a deliberately easy supervised
classification task between tori and spheres. The torus class contains closed
grids with \((h,p)\in\{(4,4),(5,5),(6,6),(7,7)\}\). The sphere class contains
an icosahedron with subdivision levels \(0,1,2,2\). Eight base complexes are
expanded into \(32\) permuted copies each, giving \(256\) samples balanced
between the two classes.

Because permutations of the same eight base complexes may occur in different
splits, this sanity experiment validates pipeline behavior on an intentionally
easy task; it is not an estimate of generalization to unseen cellulations.

The split is stratified into \(60\%\) training, \(20\%\) validation, and
\(20\%\) test data. Graph models train for \(40\) epochs, while the probes use
\(300\) full-batch epochs. Count and distance probe features are transformed
using
\[
\widetilde{x}
=\operatorname{sign}(x)\log(1+|x|)
\].

% ---------------------------------------------------------------------------
\subsection{Stratification of the Global Metric Channel
  \texorpdfstring{$C_5$}{C5}}\label{app:c5}

Reporting $C_5$ as a single status conflates objects of different
granularity. A histogram over node-to-cell distances is a coarse summary,
whereas the per-vertex profile consumed by \textsc{smcn}'s marking layer is
strictly finer; equality of the former does not imply equality of the latter.
We therefore stratify $C_5$ into seven sub-channels and compare each one
exactly in a post-hoc audit.

\begin{table}[t]
\caption{Exact post-hoc comparison of seven metric sub-channels on the
$16$ certified Fundamental Group pairs. A row is \emph{matched} only when
the corresponding statistic agrees on all $16$ pairs; otherwise it identifies
a measured leak. The aggregate channel $C_5$ remains \emph{partial} because
these seven objects do not exhaust all possible forms of metric information.}
\label{tab:c5strat}
\centering
\small
\begin{tabular}{@{}llcc@{}}
\toprule
\textbf{Sub-channel}
& \textbf{Object compared (exact multiset equality)}
& \textbf{Matched pairs}
& \textbf{Status} \\
\midrule
$C_{5.1}$
& Node-to-node shortest-path distance multiset
& $16/16$
& \Cmatch \\

$C_{5.2}$
& Eccentricity profile
& $16/16$
& \Cmatch \\

$C_{5.3}$
& Graph diameter
& $16/16$
& \Cmatch \\

$C_{5.4}$
& Aggregated node-to-cell distance histogram
& $16/16$
& \Cmatch \\

$C_{5.5}$
& Hasse-graph shortest-path distance multiset
& $\mathbf{1/16}$
& \Cleak \\

$C_{5.6}$
& Per-cell distance profile
  $\{\operatorname{sort}_{v} d(v,c)\}_{c}$
& $16/16$
& \Cmatch \\

$C_{5.7}$
& Per-vertex marking profile
  $\{\operatorname{sort}_{c} m(v,c)\}_{v}$
& $\mathbf{6/16}$
& \Cleak \\
\bottomrule
\end{tabular}

\end{table}

Here
\[
m(v,c)=\min_{u\prec c} d(v,u).
\]
The aggregated node-to-cell distance histogram $C_{5.4}$ constitutes the
information supplied to \textsc{dist\_mlp}; $C_{5.1}$--$C_{5.3}$ are additional
audited summaries not supplied to that probe. Its equality on all $16$ pairs is consistent with the
probe's $0/64$ result. By contrast, the finer marking profile $C_{5.7}$ supplied
to shortest-path \textsc{smcn} differs on $10$ of the $16$ pairs. The
Hasse-graph distance multiset $C_{5.5}$ also differs on $15$ pairs.
The $0/64$ result for \textsc{hasse\_gin} shows only that the tested Hasse
baseline did not exploit this distinction; it does not establish that the
channel is unused by every possible model.

\paragraph{The inference gap, stated explicitly.}
Failure of \textsc{dist\_mlp} does not close the finer metric channel read by
\textsc{smcn}. Table~\ref{tab:c5strat} makes this gap quantitative:
\textsc{dist\_mlp} consumes $C_{5.4}$, which is matched in the exact
post-hoc audit, whereas shortest-path \textsc{smcn} consumes $C_{5.7}$, which
differs on $10$ of the $16$ pairs. A failed probe at coarse granularity rules
out only its tested feature map, model class, and training budget; it licenses
no conclusions at finer granularity. Consequently, results obtained with
shortest-path marking are metric-confounded on those $10$ pairs and are
reported separately from the binary-marking configuration.

% ---------------------------------------------------------------------------
% ---------------------------------------------------------------------------
\subsection{Per-Pair Reporting and Seed Consistency}
\label{app:perpair}

Four seeds on one pair are repeated measurements of a single item, not four
independent samples: the complexes, certification, and words are identical, and
only initialization and batch order differ. The unit of independence is the
pair, and there are sixteen. An interval over sixty-four Bernoulli trials would
assume sixty-four independent draws and be too narrow. We report at the pair level and retain evaluation-level counts only as a secondary figure.

\begin{table}[t]\centering\small
\caption{Per-pair seed consistency, four model seeds per configuration.
Binary marking separates $13/16$ pairs under at least one seed: $11/16$
under all four seeds, $2/16$ under two seeds, and $3/16$ under no seed. Pair $7$
is separated by neither \textsc{smcn} configuration under any seed. The binary
total is $48/64$, including pair $15$ at $2/4$. The $1$-WL-bounded
column aggregates \textsc{count\_mlp}, \textsc{boundary\_mlp},
\textsc{dist\_mlp}, \textsc{gin}, \textsc{hasse\_gin} and
\textsc{homp}/\textsc{cin}, each of which separated no pair under any seed, over
$256$, $192$, $192$, $256$, $256$ and $112$ evaluations respectively. The three
word-length classes correspond exactly to the three cell-count classes.}
\label{tab:perpair}
\begin{tabular}{@{}ccccccc@{}}
\toprule
\textbf{Pair} & \shortstack{\textbf{Word}\\\textbf{len.}} &
\shortstack{\textbf{Cells}\\$(|X_0|,|X_1|,|X_2|)$} &
\shortstack{\textbf{\textsc{smcn}}\\\textbf{binary}} &
\shortstack{\textbf{\textsc{smcn}}\\\textbf{SPD}} &
\textbf{\textsc{esan}} &
\shortstack{\textbf{$1$-WL}\\\textbf{bounded}}\\
\midrule
 0 &  48 & $(58,204,144)$ & $4/4$ & $4/4$ & $3/4$ & $0/4$\\
 1 &  72 & $(82,300,216)$ & $2/4$ & $3/4$ & $2/4$ & $0/4$\\
 2 &  72 & $(82,300,216)$ & $4/4$ & $1/4$ & $4/4$ & $0/4$\\
 3 & 120 & $(130,492,360)$ & $0/4$ & $3/4$ & $1/4$ & $0/4$\\
 4 &  72 & $(82,300,216)$ & $4/4$ & $4/4$ & $4/4$ & $0/4$\\
 5 &  72 & $(82,300,216)$ & $4/4$ & $4/4$ & $4/4$ & $0/4$\\
 6 &  72 & $(82,300,216)$ & $4/4$ & $3/4$ & $0/4$ & $0/4$\\
 7 &  72 & $(82,300,216)$ & $0/4$ & $0/4$ & $0/4$ & $0/4$\\
 8 &  72 & $(82,300,216)$ & $4/4$ & $3/4$ & $3/4$ & $0/4$\\
 9 &  48 & $(58,204,144)$ & $4/4$ & $4/4$ & $4/4$ & $0/4$\\
10 &  72 & $(82,300,216)$ & $4/4$ & $4/4$ & $2/4$ & $0/4$\\
11 &  48 & $(58,204,144)$ & $4/4$ & $4/4$ & $4/4$ & $0/4$\\
12 & 120 & $(130,492,360)$ & $0/4$ & $4/4$ & $2/4$ & $0/4$\\
13 & 120 & $(130,492,360)$ & $4/4$ & $4/4$ & $0/4$ & $0/4$\\
14 &  48 & $(58,204,144)$ & $4/4$ & $4/4$ & $4/4$ & $0/4$\\
15 & 120 & $(130,492,360)$ & $2/4$ & $2/4$ & $0/4$ & $0/4$\\
\midrule
\multicolumn{3}{@{}l}{\textbf{Evaluations}} &
$48/64$ & $51/64$ & $37/64$ & $0/64$\\
\bottomrule
\end{tabular}

\end{table}

\begin{table}[t]\centering\small
\caption{Seed-consistency summary. Binary marking separates $11/16$ pairs
under every seed, $2/16$ under two seeds, and $3/16$ under no seed. All columns
use the same 16-pair denominator (Table~\ref{tab:perpair}). Every
$1$-WL-bounded baseline separates zero of sixteen pairs under every
seed.}\label{tab:consistency}
\begin{tabular}{@{}lccc@{}}
\toprule
\textbf{Consistency} & \textbf{\textsc{smcn} binary} &
\textbf{\textsc{smcn} SPD} & \textbf{\textsc{esan}}\\
\midrule
Solved by all four seeds & $11/16$ & $9/16$ & $6/16$\\
Solved by three seeds    & $0/16$ & $4/16$ & $2/16$\\
Solved by two seeds      & $2/16$ & $1/16$ & $3/16$\\
Solved by one seed       & $0/16$ & $1/16$ & $1/16$\\
Never solved             & $3/16$ & $1/16$ & $4/16$\\
\midrule
Solved by $\geq 1$ seed  & $13/16$ & $15/16$ & $12/16$\\
\bottomrule
\end{tabular}

\end{table}

\paragraph{Consistency by size class.}
\begin{center}\small
\begin{tabular}{@{}rrrcc@{}}
\toprule
Word len. & Cells & Pairs &  \textsc{smcn} SPD &  \textsc{smcn} binary\\
\midrule
$48$  & $406$ & $4$ & $4/4$  & $4/4$\\
$72$  & $598$ & $8$ & $3/8$  & $6/8$\\
$120$ & $982$ & $4$ & $2/4$  & $1/4$\\
\midrule
      &       &     & $9/16$ & $11/16$\\
\bottomrule
\end{tabular}
\end{center}

Under shortest-path marking, unanimity degrades with complex size, falling from
$4/4$ at word length $48$ to $3/8$ at $72$. Binary marking shows a different
pattern: its unanimous rate falls from $4/4$ at word length $48$ to $6/8$
at $72$ and $1/4$ at $120$. Among the pairs separated at least once, the two
least consistent under shortest-path marking are pair $2$ ($1/4$, word
length $72$) and pair $15$ ($2/4$, word length $120$), so the degradation is not
confined to a single size class. We report this rather than smoothing it away;
we cannot currently explain it.

\paragraph{What the two markings share.}
Binary and shortest-path marking both separate eight pairs unanimously,
$(0,4,5,9,10,11,13,14)$, and pair $7$ is not separated by either under any seed. Shortest-path
marking recovers two pairs that binary marking never separates,
$(3,12)$, which is the quantitative content of the metric channel's
contribution: it is concentrated on a specific minority of pairs rather than
uniformly improving the family.

% ---------------------------------------------------------------------------
\subsection{\texorpdfstring{$\pi_1$}{pi1}-Matched Control Family}
\label{app:controls}

The control pairs \emph{two members of the same $\pi_1$ class} with independent
noise realizations through the identical gate stack, plus one gate the
cross-class family does not need: \emph{distinctness}, rejecting pairs whose
words coincide up to the construction's rotation and reversal symmetry. Without
it a control could degenerate into the same-word reliability arm and sit at the
floor for a trivial reason.

\begin{table}[t]\centering\small
\caption{Separation on the cross-class family against the $\pi_1$-matched
control, with matched data, permutation, ordering, marking, and cache settings.
Permutation reliability passes on every reported cell.
$^{\dagger}$~The $56/60$ binary cross-class entry comes from a separate,
incomplete control rerun and is not the finalized $48/64$ main result in
Table~\ref{tab:brec-results}. It is reported only as an incomplete exploratory
comparison and should not be treated as a settled estimate.}\label{tab:controls}
\begin{tabular}{@{}lccc@{}}
\toprule
\textbf{Model} & \textbf{Params} &
\shortstack{\textbf{Cross-class}\\\textbf{(different $\pi_1$)}} &
\shortstack{\textbf{$\pi_1$-matched control}\\\textbf{(same $\pi_1$)}}\\
\midrule
\textsc{count\_mlp}    & $226$       & $0/64$  & $0/64$\\
\textsc{boundary\_mlp} & $226$       & $0/64$  & $0/64$\\
\textsc{dist\_mlp}     & $482$       & $0/64$  & $0/64$\\
\textsc{gin}           & $34{,}574$  & $0/64$  & $0/64$\\
\textsc{hasse\_gin}    & $34{,}638$  & $0/64$  & $0/64$\\
\textsc{homp}/\textsc{cin} & $255{,}618$ & $0/64$ & $0/64$\\
\midrule
\textsc{esan}          & $67{,}858$  & $37/64$ & $28/64$\\
\textsc{smcn}, binary  & $144{,}707$ & $56/60^{\dagger}$ & $48/64$\\
\textsc{smcn}, SPD     & $146{,}627$ & $51/64$ & $52/64$\\
\bottomrule
\end{tabular}

\end{table}

\paragraph{What the control shows.}
The control is not degenerate: every $1$-WL-bounded baseline scores $0/64$ on
both families, so the protocol does not separate arbitrary distinct
realizations. Against that floor the two \textsc{smcn} configurations behave
differently. Because the binary cross-class control rerun is incomplete,
we do not treat its $56/60$ value as directly comparable to the finalized main
result or as a settled cross-family estimate. Shortest-path marking
separates $51/64$ ($80\%$) against $52/64$ ($81\%$) --- a one-evaluation
difference rather than a meaningful observed gap.
\textsc{esan} shows a gap in the same direction as binary marking
($37/64$ against $28/64$).

The completed shortest-path comparison shows no cross-family gap. Adding the per-vertex
distance marking raises performance on the control to the cross-class level. This is consistent with the fine metric sub-channel
$C_{5.7}$ being unmatched on $10$ of $16$ pairs: the extra metric information
separates realizations irrespective of $\pi_1$. The incomplete binary
rerun is retained only as exploratory evidence; it does not establish that
separation is specific to $\pi_1$ or that the models compute $\pi_1$.

% ---------------------------------------------------------------------------
\paragraph{Zero-Message-Passing Ablation}\label{app:ej}

Because a distance marking is computed before the first learned layer, an
architecture can, in principle, separate a pair without message passing at all. We
test this directly by setting both the cellular and subcomplex message-passing
depths to zero, leaving only the embedding and pooling pipeline.

\begin{table}[H]\centering\small
\caption{Zero-message-passing ablation for shortest-path-marked SMCN. The
reported-depth solved count is reproduced from Table~\ref{tab:perpair}; its
four-seed denominator differs from the single-seed zero-layer denominator.}
\label{tab:zero-message-ablation}
\begin{tabular}{@{}lcccc@{}}
\toprule
\textbf{Configuration} & $n_{\mathrm{cin}}$ & $n_{\mathrm{sub}}$ &
\textbf{Pairs solved} & \textbf{Target gap / spread}\\
\midrule
Zero message passing & $0$ & $0$ & $\mathbf{0/16}$ & $0.19$\\
Reported depth (SPD marking) & $1$ & $2$ & $51/64$ & $27.49$\\
\bottomrule
\end{tabular}

\end{table}

\noindent
Without learned message passing, the model separates no pairs, and the
between-class gap is a fifth of the within-class spread. At the reported depth,
the gap is roughly $27$ times the spread. Separation on the Fundamental Group
family therefore requires learned message passing in this tested configuration;
the zero-layer embedding and pooling pipeline is insufficient. We note this
contrasts with the punctured
orientability pair, where a zero-layer ablation still separated the classes.

% ===========================  MC3  =========================================
\subsection{What the Certifier Retains and What the Model Receives}
\label{app:mc3}

The mathematical construction begins with an ordered signed word. The certifier
retains it; the emitted representation $\Phi(X_w)$ does not. Table~\ref{tab:phi}
makes the gap explicit because it is the precise reason a positive result
cannot be read as a recovery of the attaching map.

\begin{table}[H]\centering\small
\caption{The five quantities used to certify $\pi_1$ are all discarded by
$\Phi$. Faces reach the model as \texttt{tuple(sorted(f))}, so no cyclic order
survives; incidence matrices retain only their nonzero pattern, so no
orientation sign survives. A model separating $\Phi(X)$ from $\Phi(Y)$ therefore
demonstrates sensitivity to the \emph{supplied representation}. It does not
directly receive the attaching word, although it may exploit realization-induced
consequences of that word.}\label{tab:phi}
\begin{tabular}{@{}lcc@{}}
\toprule
\textbf{Information} & \shortstack{\textbf{Retained by}\\\textbf{certifier}} &
\shortstack{\textbf{Present in}\\$\Phi(X_w)$}\\
\midrule
Ordered attaching word $w$                        & yes & \textbf{no}\\
Generator signs ($a$ vs.\ $a^{-1}$)               & yes & \textbf{no}\\
Cyclic order of a face's boundary                 & yes & \textbf{no}\\
Orientation signs of $\partial_1,\partial_2$      & yes & \textbf{no}\\
Free reduction / $\pi_1$ certificate              & yes & \textbf{no}\\
\midrule
Unsigned incidence (nonzero pattern only)         & yes & yes\\
Rank-wise cell counts                             & yes & yes\\
Derived adjacency / coadjacency                   & yes & yes\\
Bridge indices (mediating cells)                  & yes & yes\\
Faces as unordered vertex sets                    & yes & yes\\
\bottomrule
\end{tabular}

\end{table}

\begin{table}[H]\centering\small
\caption{Controls relative to which separation on this family is measured.
The nine gated rows define the scope of the supported claim; the three audited
rows delimit it. Specifically,
beyond-$1$-WL architectures detect information beyond the matched homological,
local-incidence, vertex-link and Hasse-$1$-WL controls.}\label{tab:mc3controls}
\begin{tabular}{@{}llc@{}}
\toprule
\textbf{Control} & \textbf{Enforced by} & \textbf{Status}\\
\midrule
Algebraic $\pi_1$ certification of both members & generator gate & \Cmatch\\
Rank-wise cell counts; Euler characteristic     & generator gate & \Cmatch\\
Exact integral homology $(\mathbb Z,\mathbb Z^4,\mathbb Z)$, SNF & generator gate & \Cmatch\\
Local incidence and edge--face degree profiles  & generator gate & \Cmatch\\
Vertex-link WL profiles                         & generator gate & \Cmatch\\
Boundary structure (zero boundary edges)        & generator gate & \Cmatch\\
Word length; slot multiset; generator counts    & generator gate & \Cmatch\\
Signed exponent sums; backtrack motifs          & generator gate & \Cmatch\\
Rank-colored Hasse $1$-WL to stabilization (round $6$) & generator gate & \Cmatch\\
\midrule
Global metric, fine granularity ($C_{5.7}$)     & audited post hoc & \Cpart\\
Spectral ($C_6$)                                & audited post hoc & \Cleak\\
Cycle counts, $\ell\geq 4$ ($C_7$)              & audited post hoc & \Cpart\\
\bottomrule
\end{tabular}

\end{table}

The zero-message-passing ablation is reported once, with its numerical
reported-depth result, in Table~\ref{tab:zero-message-ablation}.

% ===========================  MC4  =========================================
\subsection{Spectral and Cycle Audits
  \texorpdfstring{($C_6$, $C_7$)}{(C6, C7)}}\label{app:mc4}

\begin{table}[t]\centering\small
\caption{Numerical post-hoc spectral audit at tolerance $10^{-8}$, together
with exact cycle and closed-walk audits, on the $16$ certified pairs. The twelve
spectral operator variants are grouped into nine rows.
No pair is cospectral for any tested operator. Girth and triangle
counts match on every pair; matching degrades sharply from $\ell=4$ and vanishes
by $\ell=7$. Exact simple-cycle enumeration costs $\approx 4\times$ per unit
length on these complexes, so $L=8$ is the enumeration limit and nothing is
claimed beyond it.}\label{tab:mc4gates}
\begin{tabular}{@{}llcc@{}}
\toprule
\textbf{Channel} & \textbf{Operator / statistic} &
\textbf{Matched pairs} & \textbf{Status}\\
\midrule
$C_6$ & $1$-skeleton adjacency $A$                     & $0/16$ & \Cleak\\
$C_6$ & $1$-skeleton Laplacian $D-A$                   & $0/16$ & \Cleak\\
$C_6$ & $1$-skeleton normalized Laplacian              & $0/16$ & \Cleak\\
$C_6$ & Hasse adjacency                                & $0/16$ & \Cleak\\
$C_6$ & Hasse Laplacian                                & $0/16$ & \Cleak\\
$C_6$ & Hasse rank-colored adjacency                  & $0/16$ & \Cleak\\
$C_6$ & Hodge $L_0$ unsigned / signed                  & $0/16$ & \Cleak\\
$C_6$ & Hodge $L_1$ unsigned / signed                  & $0/16$ & \Cleak\\
$C_6$ & Hodge $L_2$ unsigned / signed                  & $0/16$ & \Cleak\\
\midrule
$C_7$ & Girth                                          & $16/16$ & \Cmatch\\
$C_7$ & Triangles ($\ell=3$)                           & $16/16$ & \Cmatch\\
$C_7$ & Cycles $\ell=4$                                & $1/16$  & \Cleak\\
$C_7$ & Cycles $\ell=5$                                & $1/16$  & \Cleak\\
$C_7$ & Cycles $\ell=6$                                & $1/16$  & \Cleak\\
$C_7$ & Cycles $\ell=7$                                & $0/16$  & \Cleak\\
$C_7$ & Cycles $\ell=8$                                & $0/16$  & \Cleak\\
$C_7$ & Closed walks $\mathrm{tr}(A^{\ell})$, $\ell\leq 16$ & $0/16$ & \Cleak\\
\bottomrule
\end{tabular}

\end{table}

\paragraph{Relationship between $C_6$ and $C_7$.}
Closed-walk counts satisfy
$\mathrm{tr}(A^{\ell})=\sum_i\lambda_i^{\ell}$, so adjacency cospectrality would
force equality of closed-walk counts at \emph{every} length, unbounded.
Adjacency cospectrality would therefore imply equality of all closed-walk
traces; our finite simple-cycle audit remains a separate partial diagnostic.
Message passing propagates information along walks rather than directly
enumerating simple cycles, so closed-walk statistics are especially relevant
to the ladder's receptive field.

\paragraph{Observed limits of the attempted gates.}
Under the stated sampling budget, the attempted rejection sampler produced
no pair satisfying exact cospectrality in addition to the existing homology,
cell-count, word-statistic, and Hasse-WL gates. We therefore did not add a
spectral gate. Constructing cospectral realizations would require a redesigned
generator rather than the tested rejection filter.

The relevant benchmark question is whether a declared, model-accessible
channel carries class information. Table~\ref{tab:mc4control} tests that
question for the finite spectral, walk, and cycle statistics listed here.

\begin{table}[H]\centering\small
\caption{Matched-pair counts for spectral and cycle statistics on the
cross-class family and the two $\pi_1$-matched control families, using
identical gates, generator settings, and $16$ pairs per family. An entry
$x/16$ means that the stated statistic agrees for $x$ of the $16$ pairs.
}
\label{tab:mc4control}
\begin{tabular}{@{}lccc@{}}
\toprule
\textbf{Statistic} &
\shortstack{\textbf{Cross-class}\\\textbf{(different $\pi_1$)}} &
\shortstack{\textbf{surface-vs-}\\\textbf{surface}} &
\shortstack{\textbf{free-vs-}\\\textbf{free}}\\
\midrule
\multicolumn{4}{@{}l}{\emph{$C_6$ --- spectral (all twelve operators)}}\\
$1$-skeleton $A$, $D-A$, normalized Laplacian & $0/16$ & $0/16$ & $0/16$\\
Hasse adjacency, Laplacian, rank-colored     & $0/16$ & $0/16$ & $0/16$\\
Hodge $L_0,L_1,L_2$ signed and unsigned       & $0/16$ & $0/16$ & $0/16$\\
\midrule
\multicolumn{4}{@{}l}{\emph{$C_7$ --- walks and cycles}}\\
Closed walks $\mathrm{tr}(A^{\ell})$, $\ell\leq 16$ & $0/16$ & $0/16$ & $0/16$\\
Girth                                          & $16/16$ & $16/16$ & $16/16$\\
Cycles $\ell=3$ (triangles)                    & $16/16$ & $16/16$ & $16/16$\\
Cycles $\ell=4$                                & $1/16$  & $4/16$  & $9/16$\\
Cycles $\ell=5$                                & $1/16$  & $4/16$  & $9/16$\\
Cycles $\ell=6$                                & $1/16$  & $1/16$  & $4/16$\\
Cycles $\ell=7$                                & $0/16$  & $0/16$  & $0/16$\\
Cycles $\ell=8$                                & $0/16$  & $0/16$  & $0/16$\\
\bottomrule
\end{tabular}

\end{table}
\paragraph{Results}
No pair in any family is cospectral under the twelve tested operators, and
no pair has matching closed-walk counts through length $16$. Because these
differences occur at the same rate whether the two members have different or
identical fundamental groups, they are not specific to the target
$\pi_1$ contrast. In contrast, girth and triangle counts match on all pairs
and therefore provide no separation signal. For simple cycles of lengths
$4$--$6$, the $\pi_1$-matched controls agree more often than the cross-class
family, indicating a potentially class-correlated signal. At lengths $7$ and
$8$, no pair matches in any family; thus the statistic still leaks, but its
mismatch rate is not specific to the cross-class comparison. Results were
computed using \texttt{control\_gates.py}.

\paragraph{What the control establishes, and where it stops.}
For the spectral channel, the claim holds exactly as stated: no pair in any of
the three families is cospectral for any operator, so spectral leakage is a
property of the realization scheme, and members with identical $\pi_1$ leak at
the same rate as members with different $\pi_1$. The same realization-level
conclusion holds for the audited closed-walk counts. Closed walks are especially
relevant because message passing propagates information along walks rather than
directly enumerating simple cycles.
For short simple-cycle counts, the claim does not hold, and we do not make it.
At $\ell=4$--$6$, the $\pi_1$-matched families match strictly more often than the
cross-class family, so these counts are partially class-correlated rather than
purely realization noise. We therefore restrict the realization-scheme argument
to $C_6$ and to closed walks and report the short-cycle window as an
uncontrolled channel that may carry class information. Whether a model can
exploit the class-correlated mismatches at lengths $4$--$6$ is not established
here; every $1$-WL-bounded architecture in our ladder scores $0/64$
pair--seed evaluations over 16 fixed pairs.

% ===========================================================================

\section*{GenAI Usage Statement}
\phantomsection
\label{app:genai}

Generative AI tools were used to assist with language editing, manuscript
organization, code development, and the refinement of experimental and
diagnostic-control ideas. All mathematical claims, experimental designs,
implementations, results, references, and AI-assisted text were independently
reviewed and verified by the authors. The authors take full responsibility for
the final content of this work.

%% file: log_2026.bbl
\begin{thebibliography}{99}
%%%%%%%%%%%%%%%%%%%%%%%%%%%%%%%%%%%%%%%%%%%

\bibitem{ballester2024cellular}
Rub\'en Ballester, Pablo Hern\'andez-Garc\'ia, Mathilde Papillon, Claudio Battiloro, Nina Miolane, Tolga Birdal, Carles Casacuberta, Sergio Escalera, and Mustafa Hajij.
Attending to topological spaces: The cellular transformer.
\emph{arXiv preprint arXiv:2405.14094}, 2024.

\bibitem{barbarossa2020topological}
Sergio Barbarossa and Stefania Sardellitti.
Topological signal processing over simplicial complexes.
\emph{IEEE Transactions on Signal Processing}, 68:2992--3007, 2020.
doi:10.1109/TSP.2020.2981920.

\bibitem{battiloro2023attention}
Claudio Battiloro, Lucia Testa, Lorenzo Giusti, Stefania Sardellitti, Paolo Di Lorenzo, and Sergio Barbarossa.
Generalized simplicial attention neural networks.
\emph{IEEE Transactions on Signal and Information Processing over Networks}, 10:833--850, 2024.
doi:10.1109/TSIPN.2024.3485473.

\bibitem{bechlerspeicher2024graphs}
Maya Bechler-Speicher, Ido Amos, Ran Gilad-Bachrach, and Amir Globerson.
Graph neural networks use graphs when they shouldn't.
In \emph{Proceedings of the 41st International Conference on Machine Learning}, volume 235 of \emph{Proceedings of Machine Learning Research}, pages 3284--3304. PMLR, 2024.

\bibitem{bevilacqua2022esan}
Beatrice Bevilacqua, Fabrizio Frasca, Derek Lim, Balasubramaniam Srinivasan, Chen Cai, Gopinath Balamurugan, Michael M. Bronstein, and Haggai Maron.
Equivariant subgraph aggregation networks.
In \emph{Proceedings of the 10th International Conference on Learning Representations}, 2022.

\bibitem{bodnar2021cellular}
Cristian Bodnar, Fabrizio Frasca, Nina Otter, Yuguang Wang, Pietro Li\`o, Guido F. Montufar, and Michael Bronstein.
Weisfeiler and Lehman go cellular: CW networks.
In \emph{Advances in Neural Information Processing Systems}, volume 34, pages 2625--2640, 2021.

\bibitem{bodnar2021simplicial}
Cristian Bodnar, Fabrizio Frasca, Yuguang Wang, Nina Otter, Guido F. Montufar, Pietro Li\`o, and Michael Bronstein.
Weisfeiler and Lehman go topological: Message passing simplicial networks.
In \emph{Proceedings of the 38th International Conference on Machine Learning}, pages 1026--1037. PMLR, 2021.

\bibitem{carlsson2009topology}
Gunnar Carlsson.
Topology and data.
\emph{Bulletin of the American Mathematical Society}, 46(2):255--308, 2009.
doi:10.1090/S0273-0979-09-01249-X.

\bibitem{carriere2020perslay}
Mathieu Carri\`ere, Fr\'ed\'eric Chazal, Yuichi Ike, Th\'eo Lacombe, Martin Royer, and Yuhei Umeda.
PersLay: A neural network layer for persistence diagrams and new graph topological signatures.
In \emph{Proceedings of the 23rd International Conference on Artificial Intelligence and Statistics}, pages 2786--2796. PMLR, 2020.

\bibitem{dwivedi2022lspe}
Vijay Prakash Dwivedi, Anh Tuan Luu, Thomas Laurent, Yoshua Bengio, and Xavier Bresson.
Graph neural networks with learnable structural and positional representations.
In \emph{Proceedings of the 10th International Conference on Learning Representations}, 2022.

\bibitem{dwivedi2023benchmarking}
Vijay Prakash Dwivedi, Chaitanya K. Joshi, Anh Tuan Luu, Thomas Laurent, Yoshua Bengio, and Xavier Bresson.
Benchmarking graph neural networks.
\emph{Journal of Machine Learning Research}, 24(43):1--48, 2023.

\bibitem{ebli2020simplicial}
Stefania Ebli, Micha\"el Defferrard, and Gard Spreemann.
Simplicial neural networks.
In \emph{NeurIPS 2020 Workshop on Topological Data Analysis and Beyond}, 2020.

\bibitem{edelsbrunner2002persistence}
Herbert Edelsbrunner, David Letscher, and Afra Zomorodian.
Topological persistence and simplification.
\emph{Discrete \& Computational Geometry}, 28(4):511--533, 2002.
doi:10.1007/s00454-002-2885-2.

\bibitem{eitan2024blindspots}
Yam Eitan, Yoav Gelberg, Guy Bar-Shalom, Fabrizio Frasca, Michael M. Bronstein, and Haggai Maron.
Topological blindspots: Understanding and extending topological deep learning through the lens of expressivity.
In \emph{Proceedings of the 13th International Conference on Learning Representations}, 2025.

\bibitem{errica2020fair}
Federico Errica, Marco Podda, Davide Bacciu, and Alessio Micheli.
A fair comparison of graph neural networks for graph classification.
In \emph{Proceedings of the 8th International Conference on Learning Representations}, 2020.

\bibitem{geirhos2020shortcut}
Robert Geirhos, J\"orn-Henrik Jacobsen, Claudio Michaelis, Richard Zemel, Wieland Brendel, Matthias Bethge, and Felix A. Wichmann.
Shortcut learning in deep neural networks.
\emph{Nature Machine Intelligence}, 2(11):665--673, 2020.
doi:10.1038/s42256-020-00257-z.

\bibitem{giusti2023cinpp}
Lorenzo Giusti, Teodora Reu, Francesco Ceccarelli, Cristian Bodnar, and Pietro Li\`o.
CIN++: Enhancing topological message passing.
\emph{arXiv preprint arXiv:2306.03561}, 2023.

\bibitem{hajij2020cellcomplex}
Mustafa Hajij, Kyle Istvan, and Ghada Zamzmi.
Cell complex neural networks.
In \emph{NeurIPS 2020 Workshop on Topological Data Analysis and Beyond}, 2020.

\bibitem{hajij2022tdl}
Mustafa Hajij, Ghada Zamzmi, Theodore Papamarkou, Nina Miolane, Aldo Guzm\'an-S\'aenz, Karthikeyan Natesan Ramamurthy, Tolga Birdal, Tamal K. Dey, Soham Mukherjee, Shreyas N. Samaga, Neal Livesay, Robin Walters, Paul Rosen, and Michael T. Schaub.
Topological deep learning: Going beyond graph data.
\emph{arXiv preprint arXiv:2206.00606}, 2022.

\bibitem{hajij2024topox}
Mustafa Hajij, Mathilde Papillon, Florian Frantzen, Jens Agerberg, Ibrahem AlJabea, Rub\'en Ballester, Claudio Battiloro et al.
TopoX: A suite of Python packages for machine learning on topological domains.
\emph{Journal of Machine Learning Research}, 25(374):1--8, 2024.

\bibitem{hansen2019sheaf}
Jakob Hansen and Robert Ghrist.
Toward a spectral theory of cellular sheaves.
\emph{Journal of Applied and Computational Topology}, 3(4):315--358, 2019.
doi:10.1007/s41468-019-00038-7.

\bibitem{hatcher2002algebraic}
Allen Hatcher.
\emph{Algebraic Topology}.
Cambridge University Press, 2002.

\bibitem{hofer2017deep}
Christoph Hofer, Roland Kwitt, Marc Niethammer, and Andreas Uhl.
Deep learning with topological signatures.
In \emph{Advances in Neural Information Processing Systems}, volume 30, 2017.

\bibitem{horn2022togl}
Max Horn, Edward De Brouwer, Michael Moor, Yves Moreau, Bastian Rieck, and Karsten Borgwardt.
Topological graph neural networks.
In \emph{Proceedings of the 10th International Conference on Learning Representations}, 2022.

\bibitem{hotelling1931generalization}
Harold Hotelling.
The generalization of Student's ratio.
\emph{The Annals of Mathematical Statistics}, 2(3):360--378, 1931.
doi:10.1214/aoms/1177732979.

\bibitem{hu2020ogb}
Weihua Hu, Matthias Fey, Marinka Zitnik, Yuxiao Dong, Hongyu Ren, Bowen Liu, Michele Catasta, and Jure Leskovec.
Open Graph Benchmark: Datasets for machine learning on graphs.
In \emph{Advances in Neural Information Processing Systems}, volume 33, pages 22118--22133, 2020.

\bibitem{immonen2023going}
Johanna Immonen, Amauri H. Souza, and Vikas Garg.
Going beyond persistent homology using persistent homology.
In \emph{Advances in Neural Information Processing Systems}, volume 36, 2023.

\bibitem{kannan1979polynomial}
Ravindran Kannan and Achim Bachem.
Polynomial algorithms for computing the Smith and Hermite normal forms of an integer matrix.
\emph{SIAM Journal on Computing}, 8(4):499--507, 1979.
doi:10.1137/0208040.

\bibitem{kreuzer2021san}
Devin Kreuzer, Dominique Beaini, William L. Hamilton, Vincent L\'etourneau, and Prudencio Tossou.
Rethinking graph transformers with spectral attention.
In \emph{Proceedings of the 10th International Conference on Learning Representations}, 2022.

\bibitem{lim2023signnet}
Derek Lim, Joshua David Robinson, Lingxiao Zhao, Tess E. Smidt, Suvrit Sra, Haggai Maron, and Stefanie Jegelka.
Sign and basis invariant networks for spectral graph representation learning.
In \emph{International Conference on Learning Representations}, 2023.

\bibitem{lyndon1977combinatorial}
Roger C. Lyndon and Paul E. Schupp.
\emph{Combinatorial Group Theory}.
Springer, 1977.

%\bibitem{magnus1976combinatorial}
%Wilhelm Magnus, Abraham Karrass, and Donald Solitar.
%\emph{Combinatorial Group Theory: Presentations of Groups in Terms of Generators and Relations}.
%Dover Publications, 1976.

\bibitem{mahalanobis1936generalized}
Prasanta Chandra Mahalanobis.
On the generalized distance in statistics.
\emph{Proceedings of the National Institute of Sciences of India}, 2(1):49--55, 1936.

\bibitem{massey1991basic}
William S. Massey.
\emph{A Basic Course in Algebraic Topology}.
Graduate Texts in Mathematics 127. Springer, 1991.

\bibitem{morris2020tudataset}
Christopher Morris, Nils M. Kriege, Franka Bause, Kristian Kersting, Petra Mutzel, and Marion Neumann.
TUDataset: A collection of benchmark datasets for learning with graphs.
\emph{arXiv preprint arXiv:2007.08663}, 2020.

\bibitem{munkres1984elements}
James R. Munkres.
\emph{Elements of Algebraic Topology}.
Addison-Wesley, 1984.

\bibitem{papillon2023survey}
Mathilde Papillon, Sophia Sanborn, Mustafa Hajij, and Nina Miolane.
Architectures of topological deep learning: A survey of message-passing topological neural networks.
\emph{arXiv preprint arXiv:2304.10031}, 2023.

\bibitem{rampasek2022graphgps}
Ladislav Ramp\'a\v{s}ek, Michael Galkin, Vijay Prakash Dwivedi, Anh Tuan Luu, Guy Wolf, and Dominique Beaini.
Recipe for a general, powerful, scalable graph transformer.
In \emph{Advances in Neural Information Processing Systems}, volume 35, 2022.
doi:10.52202/068431-1054.

\bibitem{roddenberry2021principled}
T. Mitchell Roddenberry, Nicholas Glaze, and Santiago Segarra.
Principled simplicial neural networks for trajectory prediction.
In \emph{Proceedings of the 38th International Conference on Machine Learning}, volume 139 of \emph{Proceedings of Machine Learning Research}, pages 9020--9029. PMLR, 2021.

\bibitem{schaub2021signal}
Michael T. Schaub, Yu Zhu, Jean-Baptiste Seby, T. Mitchell Roddenberry, and Santiago Segarra.
Signal processing on higher-order networks: Livin' on the edge... and beyond.
\emph{Signal Processing}, 187:108149, 2021.
doi:10.1016/j.sigpro.2021.108149.

\bibitem{shchur2018pitfalls}
Oleksandr Shchur, Maximilian Mumme, Aleksandar Bojchevski, and Stephan G\"unnemann.
Pitfalls of graph neural network evaluation.
\emph{arXiv preprint arXiv:1811.05868}, 2018.

\bibitem{stillwell1993classical}
John Stillwell.
\emph{Classical Topology and Combinatorial Group Theory}.
Springer, 1993.

\bibitem{storjohann1996near}
Arne Storjohann.
Near optimal algorithms for computing Smith normal forms of integer matrices.
In \emph{Proceedings of the 1996 International Symposium on Symbolic and Algebraic Computation}, pages 267--274. ACM, 1996.
doi:10.1145/236869.237084.

\bibitem{telyatnikov2024topobench}
Lev Telyatnikov, Guillermo Bernardez, Marco Montagna, Mustafa Hajij, Martin Carrasco, Pavlo Vasylenko, Mathilde Papillon, Ghada Zamzmi, Michael T. Schaub et al.
TopoBench: A framework for benchmarking topological deep learning.
\emph{Journal of Data-centric Machine Learning Research}, 2025.

\bibitem{wang2024brec}
Yanbo Wang and Muhan Zhang.
An empirical study of realized GNN expressiveness.
In \emph{Proceedings of the 41st International Conference on Machine Learning}, volume 235 of \emph{Proceedings of Machine Learning Research}, pages 52134--52155. PMLR, 2024.

\bibitem{xu2019powerful}
Keyulu Xu, Weihua Hu, Jure Leskovec, and Stefanie Jegelka.
How powerful are graph neural networks?
In \emph{Proceedings of the 7th International Conference on Learning Representations}, 2019.

\bibitem{ying2021graphormer}
Chengxuan Ying, Tianle Cai, Shengjie Luo, Shuxin Zheng, Guolin Ke, Di He, Yanming Shen, and Tie-Yan Liu.
Do transformers really perform badly for graph representation?
In \emph{Advances in Neural Information Processing Systems}, volume 34, pages 28877--28888, 2021.

\bibitem{yoon2023graphworld}
Minji Yoon, Yue Wu, John Palowitch, Bryan Perozzi, and Russ Salakhutdinov.
Graph generative model for benchmarking graph neural networks.
In \emph{Proceedings of the 40th International Conference on Machine Learning}, volume 202 of \emph{Proceedings of Machine Learning Research}, pages 40175--40198. PMLR, 2023.

\bibitem{zomorodian2005computing}
Afra Zomorodian and Gunnar Carlsson.
Computing persistent homology.
\emph{Discrete \& Computational Geometry}, 33(2):249--274, 2005.
doi:10.1007/s00454-004-1146-y.

\bibitem{zopf2022wl}
Markus Zopf.
1-WL expressiveness is (almost) all you need.
In \emph{Proceedings of the 2022 International Joint Conference on Neural Networks}, IEEE, 2022.
doi:10.1109/IJCNN55064.2022.9892655.

\end{thebibliography}
